\documentclass[nofootinbib,aps,10pt]{revtex4}
\usepackage{amssymb,amsmath,graphicx,color,microtype}
\usepackage{enumerate}
\usepackage{slashed}
\usepackage{hyperref}
\usepackage{xcolor}
\usepackage{verbatim}

\hypersetup{
	colorlinks=true,
	linkcolor=blue,
	filecolor=blue,      
	urlcolor=blue, 
	bookmarks=true,
	citecolor=blue
}

\begin{document}
\title{Imprints of Schwinger Effect on Primordial Spectra} 
\author{Wan Zhen Chua$^{1}$, Qianhang Ding$^{1,2}$, Yi Wang$^{1,2}$, Siyi Zhou$^{1,2}$}  
\email{wzchua@connect.ust.hk, qdingab@connect.ust.hk, phyw@ust.hk, szhouah@connect.ust.hk}

\affiliation{${}^1$Department of Physics, The Hong Kong University of Science and Technology, \\
	Clear Water Bay, Kowloon, Hong Kong, P.R.China}
\affiliation{${}^2$Jockey Club Institute for Advanced Study, The Hong Kong University of Science and Technology, \\
	Clear Water Bay, Kowloon, Hong Kong, P.R.China} 

\begin{abstract}
We study the Schwinger effect during inflation and its imprints on the primordial power spectrum and bispectrum. The produced charged particles by Schwinger effect during inflation can leave a unique angular dependence on the primodial spectra. 
\end{abstract}

\maketitle

\section{Introduction}
Schwinger effect is a fascinating effect in quantum field theory~\cite{Schwinger:1951nm}. A pair of charged particles are produced in the vacuum, when the external electric force are strong enough. If this effect is observed, it will help our theoretical understanding of quantum field theory. But so far it has not been observed. The main obstacle is that we need $E\sim 1.3\times 10^{18} \rm V/m$~\cite{DiPiazza:2011tq}. This made us to consider Schwinger effect in astrophysical and cosmological context~\cite{Ruffini:2009hg,Tangarife:2017rgl,Tangarife:2017vnd}. In this paper, we focus on searching for the observational signature of Schwinger effect in inflation.

The existence of a large enough electric field during inflation is conventionally considered theoretically challenging. This is due to the fact that {the energy density} of radiation typically drops with the scale factor as $a^{-4}$. During inflation, the electric field and magnetic field are quickly diluted away with the rapid expansion of the universe. However, we do observe a large scale magnetic field of order micro Gauss on 10 $\rm kpc$ scale~\cite{Clarke:2000bz,Govoni:2004as,Vogt:2005xf} and $10^{-16}$ Gauss even on $\rm Mpc$ scale expected in cosmic voids~\cite{Dolag:2010ni,Neronov:1900zz,Tavecchio:2010mk}. These large scale coherent magnetic fields can hardly be explained without a primordial origin. A natural setting to generate the large scale coherent primordial magnetic field is inflation~\cite{Turner:1987bw}. However, due to the conformal invariance, the magnetic field also drops as $a^{-4}$. Lots of efforts have been made to generate the primordial magnetic field during inflation by breaking the conformal invariance~\cite{Ratra:1991bn,Dolgov:1993vg,Gasperini:1995dh,Giovannini:2000ad,Bamba:2003av,Bamba:2006ga,Giovannini:2007qn,Martin:2007ue,Subramanian:2009fu,Kandus:2010nw,Durrer:2006pc,Atmjeet:2013yta,Fujita:2015iga,Domenech:2015zzi,Stahl:2018idd}. By far, the best model we know of is still not sufficient to generate the required amount of primordial magnetic field to explain the large scale magnetic field today. One encounters the problem of either a backreaction of the electric field or a strong coupling regime at very early times \cite{Demozzi:2009fu}. This suggests that a background magnetic field  should be continuously generated during inflation to counter the effect that it is diluted away quickly. Similarly, we would expect that the same mechanism may be used to generate the electric field to compensate for the fact that the electric field is also diluted away. Such examples do exist, and they are mainly obtained by the breaking of the conformal symmetry of the gauge fields. For example, in \cite{Watanabe:2009ct,Kitamoto:2018htg}, a dilatonic coupling between the inflaton and the gauge field in the action of the type $f(\phi)^2FF$ can generate a constant electric field with energy density not changing with respect to the expansion of the universe. It is shown in \cite{Kanno:2009ei} that this constant electric field is even an attractor solution in the context of anisotropic inflation. 

In this work, we investigate the consequence that the electric field may bring us.  We {propose a simple model with} a constant electric field with an unchanged energy density in a physical volume during inflation. In short, we focus on the signatures produced. Schwinger effect is studied in  2D~\cite{Kim:2008xv,Frob:2014zka,Stahl:2015gaa} and 4D~\cite{Kobayashi:2014zza,Banyeres:2018aax,Hayashinaka:2016dnt,Geng:2017zad,Hayashinaka:2018amz} de Sitter space. Unlike the flat space case, strong electric field is not needed in inflation to produce super light particles. Charged super light particles will be mainly produced gravitationally during inflation with weak electric field. This phenomenon is known as ``hyperconductivity". 

One way to observe the Schwinger effect during inflation is to measure the properties of charged fields produced. If the charged fields are coupled to the inflaton, they will decay to inflatons during inflation, thus leaving signatures on the primordial power spectrum and bispectrum. The idea stems from the so-called quasi-single field inflation~\cite{Chen:2009we,Chen:2009zp,Baumann:2011nk}, or cosmological collider physics~\cite{Arkani-Hamed:2015bza}, which states that if there exist some massive fields of mass $m\sim H$, they can leave imprints on the squeezed limit of non-Gaussianities. Interestingly, we found that if there exist a constant electric field during inflation, the Schwinger effect will cause an angular dependence on the primordial power spectrum. This angular dependence is different from the other mechanisms that produce the angular dependence on the primordial power spectrum. For example, Bianchi universes such as anisotropic inflation generated by a vector field background \cite{Ford:1989me,Yokoyama:2008xw,Emami:2010rm,Soda:2012zm,Adshead:2018emn}, Galileons \cite{Tahara:2018orv} or higher-order of curvature terms \cite{Barrow:2005qv,Barrow:2006xb,Barrow:2009gx} can produce an angular dependence $\cos^2\theta$  (See \cite{Maleknejad:2012fw,Emami:2015qjl} for reviews and many more related works there in); inflation with a massive spin-1 field can produce the angular dependence $P_1(\cos\theta)$. Moreover, since the magnitude of non-Gaussianities is directly proportional to the number of particles produced during inflation, the bispectrum has an angular dependence as more charged particles are produced in the direction parallel with the direction of the electric field.

This paper is organized as follows, in Section~\ref{model}, we introduce the model we are considering. In Section~\ref{geodesiceq}, we derive the geodesic equation of a charged scalar particle. In Section~\ref{Powerspectrum}, we give the primordial power spectrum. In Sectrion~\ref{Bispectrum}, we give the bispectrum.  In Section~\ref{loopBispectrum}, we give the result of loop corrections to the bispectrum. We give a conclusion in Section~\ref{Conclusion}.

\section{Model}\label{model}
\begin{figure}[htbp] 
	\centering 
     \includegraphics[width=8.5cm]{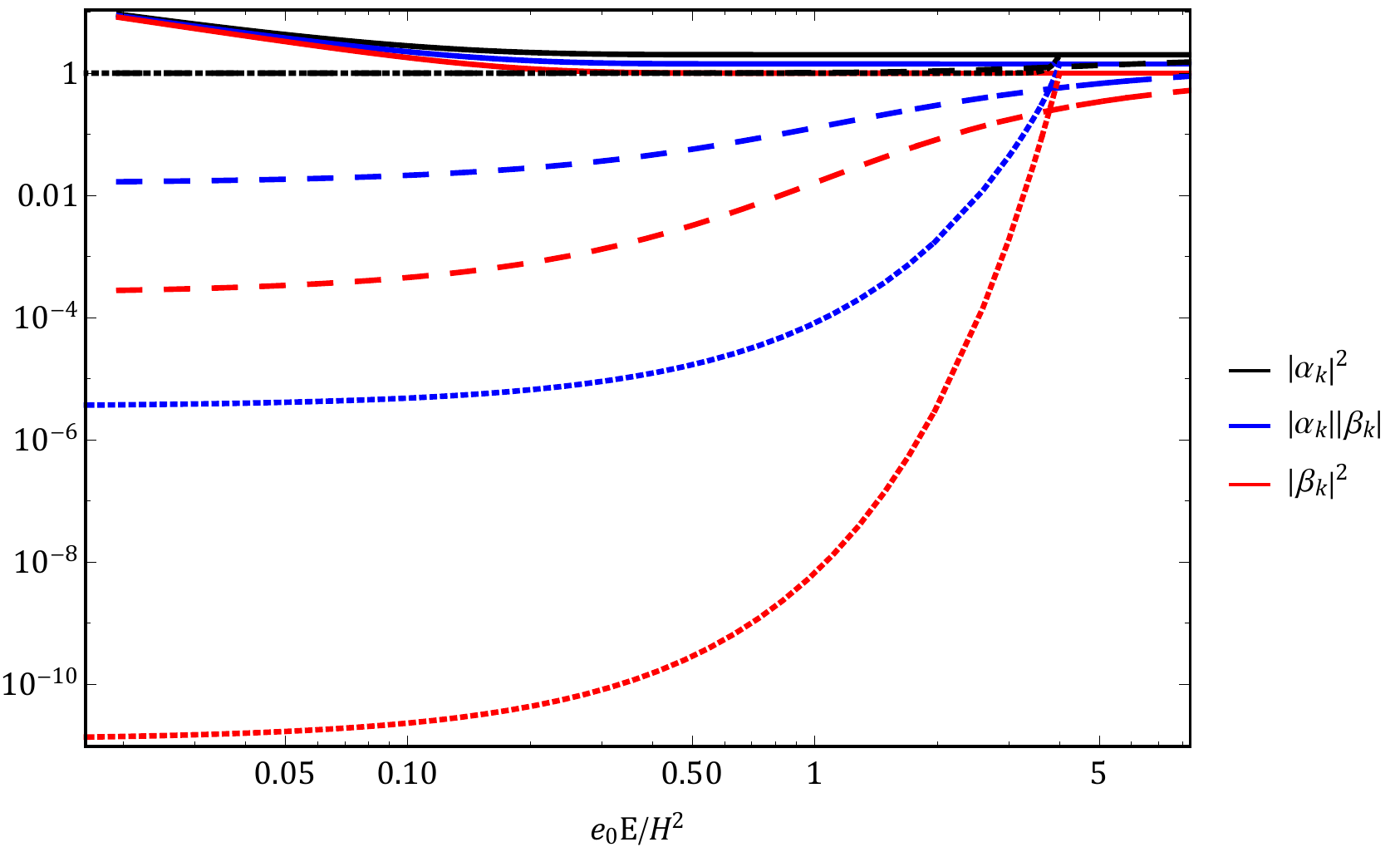}  
     \includegraphics[width=8.5cm]{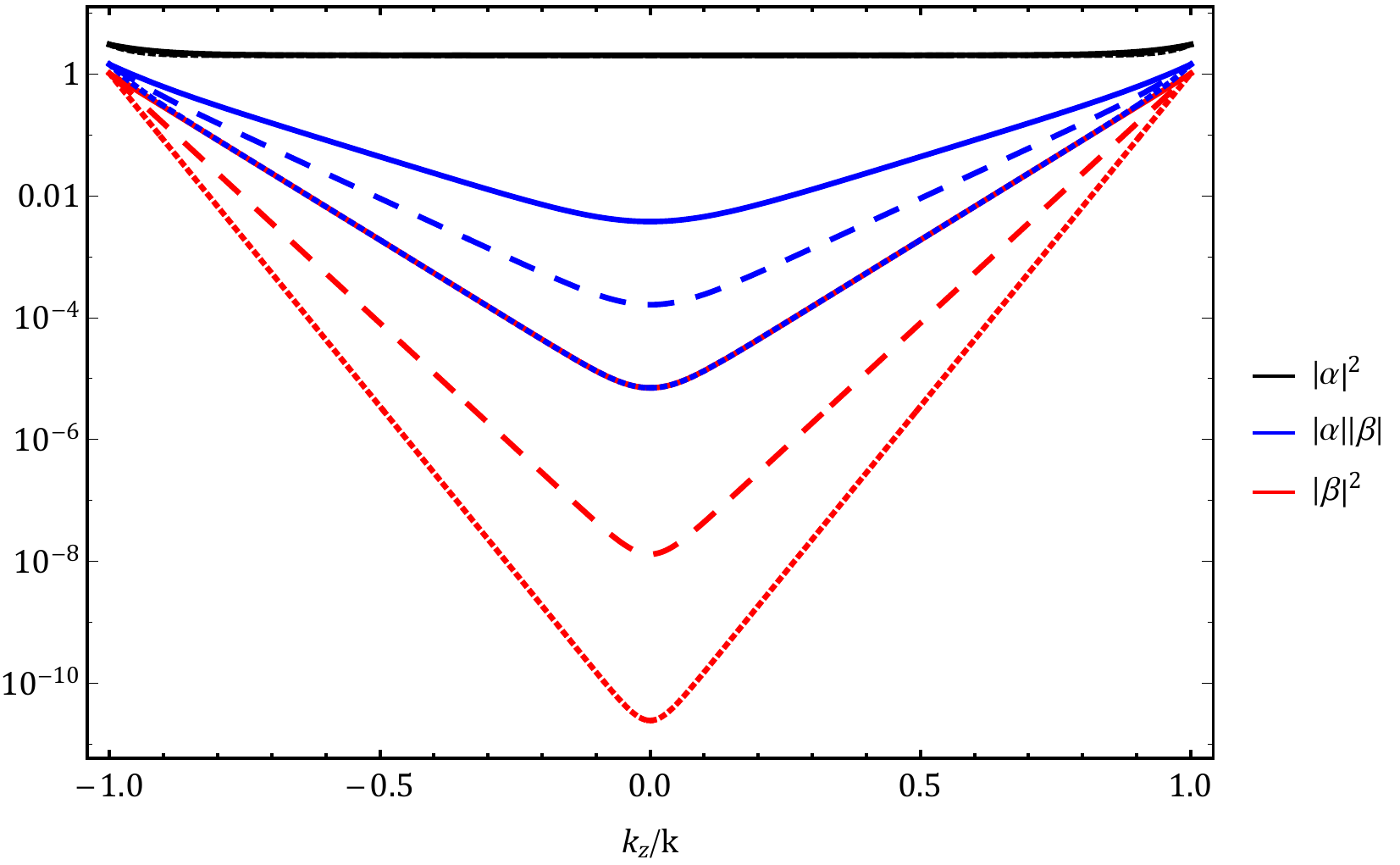}  
     \caption{The left panel of the figure shows the combinations of the Bogoliubov coefficients $|\alpha_{\mathbf k}|^2$ in black, $|\alpha_{\mathbf k}||\beta_{\mathbf k}|$ in blue and $|\beta_{\mathbf k}|^2$ in red {as a function of electric field by taking $k_z/k = 1$. For the right panel, we combine the contribution of the positive charged particles and negative charged particles to the Bogoliubov coefficients. This contributes to the symmetry of the figure.} The solid and dashed lines correspond to $m=3H/2$ and $2H$ respectively. The dotted line corresponds to fixing $|\mu|=4$ and hence the electric field $e_0E/H^2$ cannot exceed $4$ {as $\tilde{m} = \sqrt{m^2-9/4} \geq 0$}. The right panel of the figure shows the Bogoliubov coefficients as a function of $k_z/k$. The solid, dashed and dotted lines represents $\mu=2i, 3i, 4i$, respectively.}  \label{figure1}
\end{figure}

We consider {an inflation model where} QED is coupled to a pair of charged scalar $\sigma$ and $\sigma^*$ in four dimensional de Sitter space. 
\begin{align}\label{action}
S= \int d^4 x \sqrt{-g} \bigg[ -g^{\mu\nu}(\tilde{R}+\sigma)(\tilde{R}+\sigma^*)\partial_\mu \theta \partial_\nu \theta- g^{\mu\nu} D_\mu \sigma^* D_\nu \sigma - V_{\text{sr}}(\theta)- m^2|\sigma|^2 -\frac{1}{4}f(\theta)^2 F_{\mu\nu}F^{\mu\nu}  \bigg]~,
\end{align} 
where {$D_\mu \equiv \partial_\mu + i e(\theta_0) A_\mu$}. $F_{\mu\nu} = \partial_{\mu} A_\nu - \partial_\nu A_{\mu} $ is the electromagnetic tensor. Note that this action has a nonzero curvature in the field space. The curvature of the field space is zero if we use the Cartesian coordinate, while it becomes nonzero if we use the polar coordinate and make the radial coordinate complex ($\sigma$ in this case). We cannot make the radial coordinate complex without introducing curved field spaces. The FRW metric is
\begin{align}\label{metric}
	ds^2 = a^2 (\tau) (-d \tau^2+ d\mathbf x^2)~,
\end{align}
where $\tau$ is the conformal time. We {study} effects of backreaction {in} Appendix \ref{backreaction}. We consider a constant electric force in the $z$ direction,
\begin{align}\label{gauge_field}
e(\theta_0)A_\mu = \frac{Ee_0}{H^2 \tau} \delta^z_\mu, \quad E = {\rm const}~,
\end{align}
{where $e_0$ is a constant defined in Appendix \ref{backreaction}.} Thus, the equation of motion for the $\sigma$ field {in the large $\tilde{R}$ limit is (For the Lagrangian of the free $\delta\sigma$ field, please refer to Appendix \ref{3rdlagrangian})}
\begin{align}\label{sigma_eqm}
\delta\sigma'' + 2 \frac{a'}{a} \delta\sigma' - \partial_i \partial_i \delta\sigma - 2 i e(\theta_0) A_z \partial_z \delta\sigma + e(\theta_0)^2 A_z^2 \delta\sigma + a^2 m^2 \delta\sigma = 0~.
\end{align}
We quantize the $\delta\sigma$ field in the following way
\begin{align}
	\delta\sigma_{\mathbf k} & = v_{\mathbf k} a_{\mathbf k} + v_{\mathbf k}^* b^\dagger_{-\mathbf k}~, \\
	\delta\sigma^*_{\mathbf k} & = v_{-\mathbf k}^* a^\dagger_{-\mathbf k} + v_{-\mathbf k} b_{\mathbf k}~,
\end{align}
where $a_{\mathbf k}$ and $a^\dagger_{\mathbf k}$ are annihilation and creation operators of the positively charged scalar particle, and $b_{\mathbf k}$ and $b^\dagger_{\mathbf k}$ are annihilation and creation operators of the negatively charged scalar particle. They satisfy the commutation relations
\begin{align}
	 & [a_{\mathbf k}, a^\dagger_{\mathbf p}] = [b_{\mathbf k}, b^\dagger_{\mathbf p}] = (2\pi)^3 \delta^{(3)} (\mathbf k - \mathbf p)~, \\ 
	 & [a_{\mathbf k}, a_{\mathbf p} ] = [b_{\mathbf k}, b_{\mathbf p} ] = [a_{\mathbf k}, b_{\mathbf p} ] = [a_{\mathbf k}, b^\dagger_{\mathbf p} ] = \cdots = 0 ~.
\end{align} 
We introduce the variables
\begin{align}\label{parameters}
z \equiv 2 k i \tau,\quad  \kappa \equiv - i  \frac{k_z}{k} \frac{e_0E}{H^2}, \quad \mu^2 \equiv \frac{9}{4} - \frac{e_0^2 E^2}{H^4} - \frac{m^2}{H^2}~,
\end{align}
where $\kappa$ is imaginary. The real part of the parameter $\kappa$ characterizes the magnitude of the electric field projected to the direction of the trajectory of the negative charged particle. In this work, we focus on the parameter regime where $e_0^2E^2/H^4+m^2/H^2>9/4$, thus $\mu$ is imaginary. Our work can be easily generalized to the $e_0^2E^2/H^4+m^2/H^2<9/4$ case. The real part of the parameter $\mu$ can be understood as the effective mass of a charged particle in de Sitter space in Hubble units with correction $9/4$ coming from the curved space time and $e_0^2E^2/H^2$ from the electric field. The mode function satisfies the equation
\begin{align}
	\frac{d^2}{dz^2} (a v_{\mathbf k}) + \bigg\{ \frac{1}{z^2} \bigg( \frac{1}{4} - \mu^2 \bigg) + \frac{\kappa}{z} - \frac{1}{4} \bigg\} (a v_{\mathbf k}) = 0~.
\end{align}
There are two solutions, which are given by the Whittaker functions $W_{\kappa,\mu}(z)$ and $M_{\kappa,\mu}(z)$. Since in the sub-horizon limit $|z|\rightarrow\infty$, the solution must approach to the Minkowski solution, we obtain the mode function
\begin{align}\label{solution}
	a v_{\mathbf k} = \frac{e^{i \kappa \pi/2}}{\sqrt{2 k}} W_{\kappa,\mu} (z)~.
\end{align}
In the late time limit, the mode function behaves as
\begin{align}\label{latetimebehavior}
	a v_{\mathbf k} = \frac{e^{-|\mu|\pi/2}}{2\sqrt{k|\mu|}} \bigg\{ \alpha_{\mathbf k} M_{\kappa,\mu} (z) + \beta_{\mathbf k} (M_{\kappa,\mu}(z))^* \bigg\}~.
\end{align}
The coefficients $\alpha_{\mathbf k}$ and $\beta_{\mathbf k}$ satisfies the normalization condition
\begin{align}
	|\alpha_{\mathbf k}|^2 - |\beta_{\mathbf k}|^2 = 1~.
\end{align}
The Bogoliubov coefficients can be obtained as 
\begin{align}
	\alpha_{\mathbf k} = (2|\mu|)^{1/2} e^{(i\kappa+|\mu|)\pi/2} \frac{\Gamma(-2\mu)}{\Gamma(\frac{1}{2}-\mu-\kappa)}, \quad \beta_{\mathbf k} = - i (2|\mu|)^{1/2} e^{(i\kappa-|\mu|)\pi/2}\frac{\Gamma(2\mu)}{\Gamma(\frac{1}{2} +\mu-\kappa)}~.
\end{align}
The qualitative feature of $|\alpha_{\mathbf k}|^2$, $|\alpha_{\mathbf k}||\beta_{\mathbf k}|$, $|\beta_{\mathbf k}|^2$ are plotted in FIG.~\ref{figure1}. We see that as $E$ increases, both $|\alpha_{\mathbf k}|^2$ and $|\beta_{\mathbf k}|^2$ first decrease exponentially and then eventually approach to a constant value, with $|\alpha_{\mathbf k}|^2$ approaching to $2$ and $|\beta_{\mathbf k}|^2$ approaching to $1$.  The number of charged particles being produced with charge {$e_0$} and wave number $\mathbf k$ per comoving three volume $\int d^3 k/(2\pi)^3$ is
\begin{align}
	n_{\mathbf k} = |\beta_{\mathbf k}|^2  = \frac{e^{2 i \kappa \pi}+e^{-2|\mu|\pi}}{2 \sinh(2|\mu|\pi)}~.
\end{align}
In the classical limit, the particle production rate is given approximately by
\begin{align}\label{pairproductionrate} 
	n_{\mathbf k}&\approx e^{-S_E} = e^{-2\pi(|\mu| \pm |\kappa|)} = e^{-S_{\pm}} = e^{-2\pi\frac{\tilde{m}}{H}(\sqrt{1+l^2}\pm \frac{|k_{z}|}{k}l)}~,
\end{align} 
where {$l\equiv e_0E/\tilde{m}H $ }characterizes the relative magnitude of the electric field and mass. $\tilde{m}^2 = m^2 - 9 H^{2}/4$ is the effective mass of a neutral particle in de Sitter space. $S_+$ is the action corresponding to the action of the process that the charged particles are produced but moving to the direction that increases the electric potential energy of itself, whereas $S_-$ corresponds to the action of the process that the charged particles are produced and moving to the direction that decreases the electric potential energy. It is always the $S_-$ that gives the dominant contribution.

There are two interesting limits that we can discuss this problem quite intuitively. The first limit is the weak electric field limit, where $l\ll1$. The classical actions $S_{\pm}$ in \eqref{pairproductionrate} can be approximated as
\begin{align}  
	S_{\pm}=2 \pi\bigg(\frac{\tilde{m}}{H}\pm\frac{|k_{z}|}{k}\frac{e_0E}{H^2}\bigg)~.
\end{align}
The first term can be understood as the usual Boltzman factor coming from the production of neutral massive particles of effective mass $\tilde m$. From the point of view of a geodesic observer, de Sitter space is associated with a thermal bath with the Hawking temperature $T=H/(2\pi)$. The second term is understood as the chemical potential from the electric field. This chemical potential can assist the production of charged particles along the direction of decreasing potential. Since in this limit, the electric field is very weak, the dominant contribution comes from the first term. Hence, although the particles are charged, they are mainly produced gravitationally due to the expansion of the universe. In~\cite{Adshead:2015kza,Adshead:2018oaa,Chen:2018xck}, other examples of the chemical potential is also discussed in the context of fermion production in inflation.

The other limit is the large electric field limit, where $l\gg1$. In this limit, the electric field is so strong that the modes do not feel the curvature of the spacetime. The classical actions $S_{\pm}$ can be approximated as
\begin{align} 
	S_{+} = 2\pi \frac{e_0E}{H^2}\bigg(1+\frac{|k_{z}|}{k}\bigg)\quad S_{-} = 2\pi\bigg(\frac{e_0E}{H^2}\bigg(1-\frac{|k_{z}|}{k}\bigg)+\frac{\tilde{m}^2}{2e_0E}\bigg)~.
\end{align}
If we consider the charged particle pairs moving along the $z$ direction, the second term dominates. As we can see, it also reproduces the flat spacetime result. 

In order to study the time scale of mass production of charged particles, it is useful to consider the WKB approximation of solution \eqref{solution}. 
\begin{align}
	a v_{\mathbf k} = \frac{1}{\sqrt{2|w_{\mathbf k}|}} \exp\bigg\{ - i \int^\tau d\tau |w_{\mathbf k}| \bigg\}~, 
\end{align}
where $w_{\mathbf k}$ is the effective frequency given by
\begin{align}\nonumber
w_{\mathbf k}^2 & = (k_z+e(\theta_0)A_z)^2+k_x^2+k_y^2+a^2m^2-\frac{a''}{a} \\
	& = \frac{1}{\tau^2} \bigg(\frac{e_0^2E^2}{H^4} + \frac{m^2}{H^2} - 2\bigg) + \frac{2}{\tau} \frac{k_z e_0E}{H^2} + k^2~, \quad k\equiv(k_x^2+k_y^2+k_z^2)^{1/2}~.
\end{align}
Then the adiabatic parameter is evaluated as
\begin{align}
\bigg| \frac{ w'_{\mathbf k}}{w_{\mathbf k}^2} \bigg| = \bigg| \frac{ H^2 (e_0^2 E^2+  e_0 E  H^2 k_z \tau -2 H^4+H^2 m^2)}{ \left( e_0^2E^2+2  e_0E H^2 k_z \tau -2 H^4+H^2 m^2 +k^2 H^4\tau^2  \right)^{3/2}} \bigg|~.
\end{align}
It is around the time 
\begin{align}
	\tau \sim -\frac{1}{k} \bigg(|\mu|^2+\frac{1}{4}\bigg)^{1/4}~,
\end{align}
that the quantity $w'_{\mathbf k}/w_{\mathbf k}^2$ approaches its maximum. This means that most particles are produced at this time scale. 

Now we observe the production of the particle via the Schwinger effect during inflation. One may think about the mechanism in the context of quasi-single field inflation. The charged particles can decay into the primordial curvature perturbations, thus leaving an imprint on the primordial power spectrum and bispectrum.
$\zeta$ is the primordial curvature perturbation. The second order action of the primordial curvature perturbation can be written down following the procedure in~\cite{Chen:2010xka,Wang:2013eqj}
\begin{align}
S_\zeta = M_p^2 \int d t \frac{d^3 k}{(2\pi)^3} \epsilon (a^3 \dot\zeta^2 - k^2 a \zeta^2)~.
\end{align}
Quantizing it in the following way
\begin{align}
\zeta_{\mathbf k} & = u_{\mathbf k} c_{\mathbf k} + u_{\mathbf k}^* c^\dagger_{-\mathbf k}  ~,
\end{align}
where $c^\dagger_{\mathbf k}$,  $c_{\mathbf k}$ are the creation and annihilation operators satisfying
the usual commutation relations
\begin{align}
& [c_{\mathbf k}, c^\dagger_{\mathbf p}] = (2\pi)^3 \delta^{(3)} (\mathbf k - \mathbf p)  ~.
\end{align}
The mode function satisfies the following equation of motion
\begin{align}
	\ddot \zeta +(3+\eta) H \dot \zeta +\frac{k^2}{a^2} \zeta = 0~.
\end{align}
To the lowest order in slow roll parameter, the solution is
\begin{align}
u_{\mathbf k} (\tau) = \frac{H}{2\sqrt{\epsilon} M_{\rm pl}} \frac{1}{k^{3/2} } (1+i k \tau) e^{-i k \tau}~.
\end{align}
We consider the following coupling between the primordial curvature perturbation and the positive charged scalar fields
\begin{align}\label{interactions1}
L_{\delta\sigma \zeta^\prime} &= c_2 \int d^3 x d \tau a^3 \delta\sigma \zeta^\prime, \quad L_{\delta\sigma \zeta^\prime \zeta^\prime} = c_3 \int d^3 x d\tau a^2 \delta\sigma \zeta^\prime \zeta^\prime~,\\
 c_2 &  = -\frac{ 2 R \dot{\theta}_0^2  }{ H} , \quad c_3 = \frac{\dot{\theta}^2_0 R}{H^2}~.
\end{align} 
The coupling between the inflaton and the negative charged scalar fields are
\begin{align}\label{interactions2}
L_{\delta\sigma^* \zeta^\prime} &= c^*_2 \int d^3 x d \tau a^3 \delta\sigma^* \zeta^\prime, \quad L_{\delta\sigma^* \zeta^\prime \zeta^\prime} = c_3^* \int d^3 x d\tau a^2 \delta\sigma^* \zeta^\prime \zeta^\prime~,\\
 c_2^* &  = -\frac{ 2 R \dot{\theta}_0^2  }{ H}, \quad c_3^* = \frac{\dot{\theta}^2_0 R}{H^2}~,
\end{align}
and the coupling between the primordial curvature perturbation and the positive and negative charged scalar fields are
\begin{align}\label{interactions3}
L_{ \delta\sigma \delta\sigma^* \zeta^\prime} &= c^\prime_2 \int d^3 x d \tau a^3 \delta\sigma \delta\sigma^* \zeta^\prime, \quad L_{\delta\sigma \delta\sigma^* \zeta^\prime \zeta^\prime} = c^\prime_3 \int d^3 x d \tau a^2 \delta\sigma \delta\sigma^* \zeta^\prime \zeta^\prime~,\\
 c_2' &  = -\frac{\dot{\theta}_0^2(\sigma_0+R)}{H \sigma_0}, \quad
c_3' = \frac{ \dot{\theta}_0^2 }{H^2}~,
\end{align} 
where $c_2$, $c_3$, $c_2^*$, $c_3^*$, $c^\prime_2$ and $c^\prime_3$ are some constants. { Here we pick some of the possible interacting terms  in \eqref{interactions1}, \eqref{interactions2} and \eqref{interactions3} (For the full analysis of the Lagrangian, please refer to Appendix \ref{3rdlagrangian}). For $c_3$, $c_3^*$, $c^\prime_2$ and $c^\prime_3$, we only show the leading order term involved. }We can also see that \eqref{interactions1} and \eqref{interactions2} do not conserve the charge of $\delta\sigma$. They correspond to cases where the phase symmetry of $\delta\sigma$ is broken, for example, in an Abelian Higgs model (see \cite{Kumar:2017ecc} for discussion of a similar case). In other words, the gauge invariance is broken. The breaking of U(1) gauge invariance can have some benefits. For example, the strong coupling problem and backreaction can be evaded \cite{Domenech:2015zzi}, though the mechanism to obtain it is still an open problem. In this case, tree level contribution dominates correction to the spectra. Equation \eqref{interactions3} corresponds to the case where charge is conserved. In this case, loop diagrams has to be computed. One may worry that these background values contribute to the mass of the gauge field and it won't be able to support a long range force. But now we are considering very small coefficients. In order to calculate the primordial spectrums, we used the Schwinger-Keldysh formalism~(For the application in quasi-single field inflation, see \cite{Chen:2017ryl}). Now we derive the four types of free propagators for both the curvature perturbation and the massive charged scalar fields. For the curvature perturbation sector, the generating functional can be written as
\begin{align}
	Z_0 [J_+,J_-] \equiv \int  \mathcal{D} \mathcal \zeta_+ \mathcal{D}\mathcal \zeta_- \exp \bigg[i \int_{\tau_0}^{\tau_f} d\tau d^3 x \bigg(\mathcal L_0[\zeta_+] -\mathcal L_0 [\zeta_-] + J_+ \zeta_+ - J_-\zeta_- \bigg) \bigg]~.
\end{align}
The four propagators can be generated using 
\begin{align}
	-i\Delta_{ab}(\tau_1,\mathbf x_1;\tau_2,\mathbf x_2) = \frac{\delta}{i a \delta J_a(\tau_1,\mathbf x_1)} \frac{\delta}{i b \delta J_b(\tau_2,\mathbf x_2)} Z_0[J_+,J_-] \bigg|_{J_{\pm}=0}~, \quad a,b=\pm~.
\end{align}
Fourier transforming it into momentum space gives~
\begin{align}
	G_{ab}({\mathbf k} ;\tau_1,\tau_2) =-  \int d^3 \mathbf x e^{-i \mathbf k\cdot \mathbf x} \Delta_{ab} (\tau_1,\mathbf x;\tau_2, \mathbf 0)~.
\end{align}
The four types of propagators are given as the following
\begin{eqnarray}\label{propagator1z}
G_{++}({\mathbf k} ,\tau_1,\tau_2)&=&\theta(\tau_1-\tau_2) u_{\mathbf k} (\tau_1)  u_{\mathbf k}^* (\tau_2)+\theta(\tau_2-\tau_1) u_{\mathbf k}^* (\tau_1) u_{\mathbf k} (\tau_2) \nonumber\\
G_{+-}({\mathbf k} ,\tau_1,\tau_2)&=& u_{\mathbf k}^* (\tau_1) u_{\mathbf k} (\tau_2)\label{propagator2z}\nonumber\\
G_{-+}({\mathbf k} ,\tau_1,\tau_2)&=& u_{\mathbf k} (\tau_1) u_{\mathbf k}^* (\tau_2)\label{propagator3z}\nonumber\\
G_{--}({\mathbf k} ,\tau_1,\tau_2)&=&\theta(\tau_1-\tau_2) u_{\mathbf k}^* (\tau_1) u_{\mathbf k} (\tau_2)+\theta(\tau_2-\tau_1) u_{\mathbf k} (\tau_1) u_{\mathbf k}^* (\tau_2)\label{propagator4z}~.
\end{eqnarray}
For charged massive scalar pairs, we need to introduce two more sources $J^*_+$ and $J^*_-$ to source the complex conjugate of the $\sigma$ field
\begin{align}\nonumber
&Z_0 [J_+,J_-,J^*_+,J^*_-] \\
&\equiv \int \mathcal  D\delta\sigma_+ \mathcal D\delta\sigma_- \mathcal D\delta\sigma^*_+ \mathcal D\delta\sigma^*_- \exp \bigg[i \int_{\tau_0}^{\tau_f} d\tau d^3 x \bigg(\mathcal L_0[\delta\sigma_+,\delta\sigma^*_+] -\mathcal L_0 [\delta\sigma_-,\delta\sigma_-^*] + J_+ \delta\sigma_+ - J_-\delta\sigma_-  + J^*_+ \delta\sigma^*_+ - J^*_-\delta\sigma^*_- \bigg) \bigg]~.
\end{align}
The four propagators can be generated using 
\begin{align}
-i\Delta_{ab}(\tau_1,\mathbf x_1;\tau_2,\mathbf x_2) = \frac{\delta}{i a \delta J_a(\tau_1,\mathbf x_1)} \frac{\delta}{i b \delta J^*_b(\tau_2,\mathbf x_2)} Z_0[J_+,J_-,J^*_+,J^*_-] \bigg|_{J_{\pm}=0,J^*_{\pm}=0}~, \quad a,b=\pm~.
\end{align}
Then we have
\begin{align}
\begin{pmatrix}
D_{++}(\mathbf k, \tau; \mathbf k' , \tau'  ) & D_{+-} (\mathbf k, \tau; \mathbf k' , \tau'  )\\
D_{-+} (\mathbf k, \tau; \mathbf k' , \tau'  ) & D_{--} (\mathbf k, \tau; \mathbf k' , \tau'  )
\end{pmatrix} = 
i\begin{pmatrix}
\langle T \delta\sigma_{\mathbf k} (\tau) \delta\sigma^{*}_{\mathbf k'} (\tau') \rangle & \langle  \delta\sigma^{*}_{\mathbf k} (\tau) \delta\sigma_{\mathbf k'} (\tau') \rangle \\
\langle  \delta\sigma_{\mathbf k} (\tau) \delta\sigma^{*}_{\mathbf k'} (\tau') \rangle & \langle \bar{T} \delta\sigma^{*}_{\mathbf k} (\tau) \delta\sigma_{\mathbf k'} (\tau') \rangle
\end{pmatrix}~.
\end{align}
The four types of propagators in the Schwinger-Keldysh formalism are 
\begin{eqnarray}\label{propagator1}
D_{++}({\mathbf k} ,\tau_1,\tau_2)&=&\theta(\tau_1-\tau_2) v_{\mathbf k} (\tau_1)  v_{\mathbf k}^* (\tau_2)+\theta(\tau_2-\tau_1) v_{\mathbf k}^* (\tau_1) v_{\mathbf k} (\tau_2)\nonumber \\
D_{+-}({\mathbf k} ,\tau_1,\tau_2)&=& v_{\mathbf k}^* (\tau_1) v_{\mathbf k} (\tau_2)\label{propagator2}\nonumber\\
D_{-+}({\mathbf k} ,\tau_1,\tau_2)&=& v_{\mathbf k} (\tau_1) v_{\mathbf k}^* (\tau_2)\label{propagator3}\nonumber\\
D_{--}({\mathbf k} ,\tau_1,\tau_2)&=&\theta(\tau_1-\tau_2) v_{\mathbf k}^* (\tau_1)  v_{\mathbf k} (\tau_2)+\theta(\tau_2-\tau_1) v_{\mathbf k} (\tau_1) v_{\mathbf k}^* (\tau_2)\label{propagator4}~.
\end{eqnarray}
 
\section{The Geodesic Equation}\label{geodesiceq}
To understand the charged particles motion in inflation, we solve the geodesic equation as an intuitive understanding. 
Following from our metric in (\ref{metric}), the following geodesic equation for a massive charged particle can be written down following the standard procedure (see text books \cite{WeinbergGravitationandCosmology,Carroll:2004st}). 
\begin{align}
	\frac{d^2 x^\mu}{ds^2} = - \Gamma^\mu_{\alpha\beta} \frac{dx^\alpha}{ds} \frac{dx^\beta}{ds} + \frac{e(\theta_0)}{m} F^{\mu\beta} \frac{dx^\alpha}{ds} g_{\alpha\beta}, \quad
g_{\alpha\beta} \frac{dx^\alpha}{ds} \frac{dx^\beta}{ds} = -1~,
\end{align}
where the connection is
\begin{align}
\Gamma^\lambda_{\alpha\beta} = \frac{1}{2} g^{\lambda\tau} \bigg(\frac{\partial g_{\tau\alpha}}{d x^\beta} + \frac{\partial g_{\tau\beta}}{\partial x^\alpha} - \frac{\partial g_{\alpha \beta}}{\partial x^\tau}\bigg)~.
\end{align} 
Here, we would like to observe the change in the physical velocity of the massive charged particle. We would like to solve the following geodesic equation:
\begin{align}
\frac{d^2x(t)}{dt^2} =  -2 \frac{\dot{a}}{a} \frac{dx(t)}{dt} + \frac{e_0}{m}a^{-1}E~.
\end{align}
The initial condition we would choose is $\dot{x}(t_0)=0$, which means that the particles are produced at zero velocity. The solution to this equation subjected to the initial condition is
\begin{align}
	x(t) = \frac{E e_0}{m H^2}e^{-H t} \bigg(\text{cosh}(H(t-t_0))-1 \bigg)~.
\end{align} 
The velocity of the particle is
\begin{align}
	\dot{x}(t) = \frac{E e_0}{m H} \bigg(e^{-H t} - e^{H t_0-2 H t}\bigg)~.
\end{align}
At first, the force from electric field dominates over the Hubble friction and the particle starts to accelerate. Later, the particle decelerates due to the Hubble friction. The maximum velocity occurs in the subhorizon. 
\begin{figure}[htbp] 
	\centering 
	\includegraphics[width=12cm]{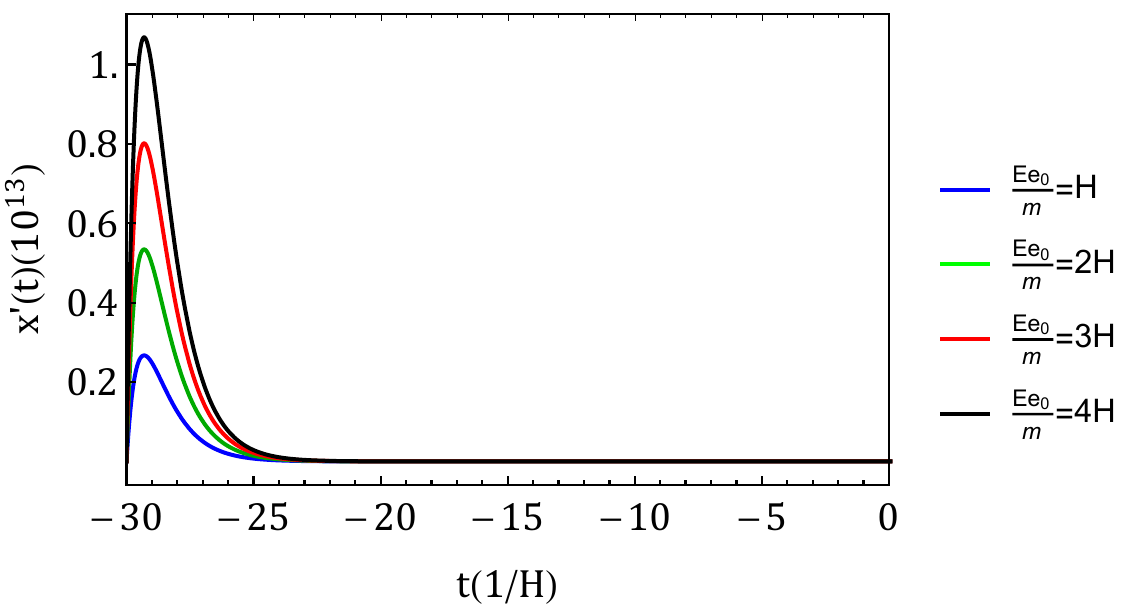}   
	\caption{The comoving velocity of the particle with respect to the physical time $t$. We take the Hubble parameter to be $1$ in making the plot. We choose the initial time to be $t_0=-30$ and $t_f = 30$ to ensure 60 e-folds. We here plot the evolution of the velocity in the subhorizon level.  Initially, the velocity of the particle is zero. Through the acceleration from the electric field, the velocity of the particle increases and the direction depends on the charge of the particle. For stronger electric fields,the maximum velocity attained is much larger. After some time, the particle starts to decelerate and its velocity slowly decreases to zero due to the Hubble expansion of the universe. We can see the velocity has decreased to zero at subhorizon level. Hence, the change in the frequency of the oscillatory signal in the bispectrum can't be observed as the bispectrum is imprinted at late times. } \label{geodesic}
\end{figure}
 
\section{Power Spectrum}\label{Powerspectrum}
\begin{figure}[htbp] 
	\centering 
	\includegraphics[width=4.5cm]{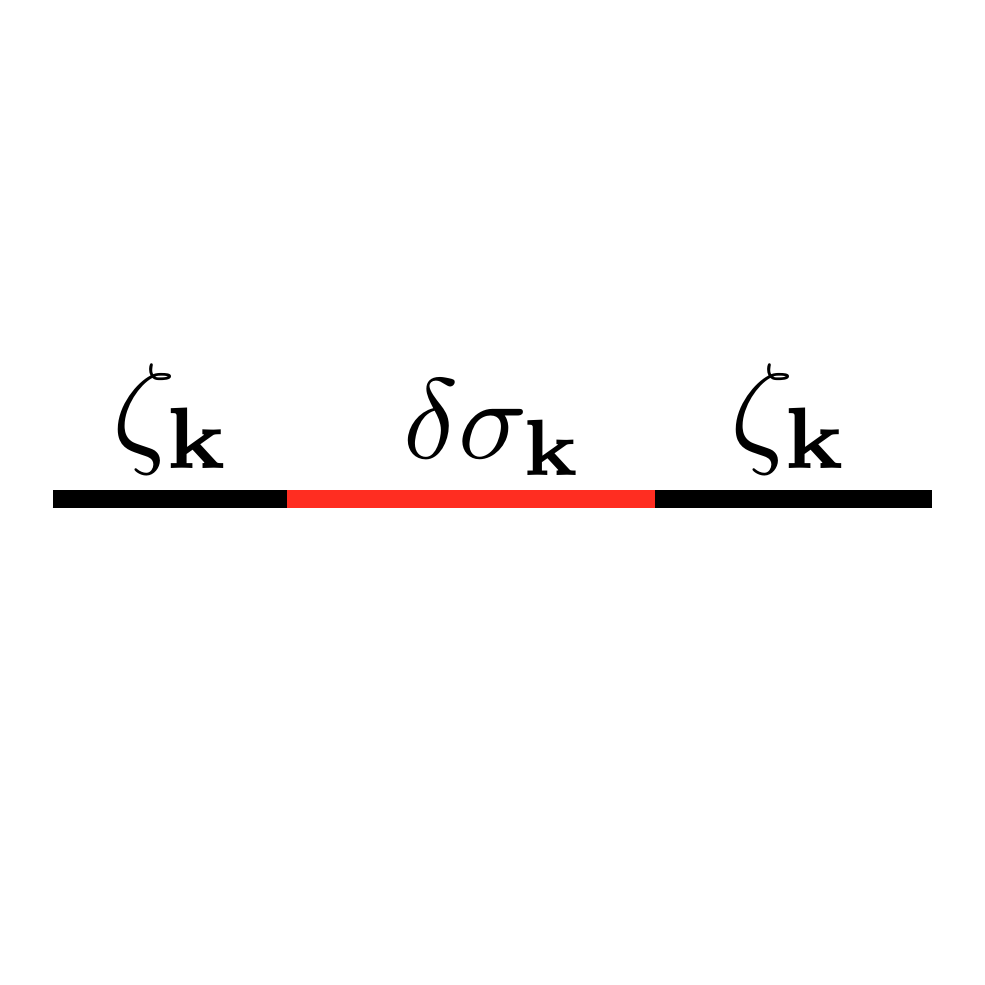}   
	\caption{{The Feynman diagram we are considering for the correction to the power spectrum. This diagram involves the interaction of $\delta\sigma$ with the inflaton. For the total correction, we need to include the correction from $\delta\sigma^*$.}} \label{fenman2}
\end{figure}

In this section, we study the {correction to }the power spectrum of our model, which can be shown in FIG. \ref{fenman2}. It can be evaluated using the Schwinger-Keldysh formalism \cite{Chen:2017ryl}
\begin{align}\label{powerspectrum1}
\langle \zeta_{\mathbf k} \zeta_{-\mathbf k} \rangle'  =  |c_2|^2 \sum_{a,b = \pm} ab \int_{-\infty}^{0}   \int_{-\infty}^{0}   \frac{d\tau_1}{(-H\tau_1)^3} \frac{d\tau_2}{(-H\tau_2)^3} & \partial_{\tau_1}G_{a+}({\mathbf k} ,\tau_1,0) D_{ab}({\mathbf k} ,\tau_1,\tau_2) \partial_{\tau_2}G_{b+}({\mathbf k} ,\tau_2,0) + ({\kappa\rightarrow-\kappa})~,
\end{align}
where $G(\mathbf{k},\tau_1,\tau_2)$ and $D(\mathbf{k},\tau_1,\tau_2)$ are defined by (\ref{propagator1z}) and (\ref{propagator1}) respectively. Here, we need to do a sum over of all the $+$ and $-$ modes. The $'$ means the momentum conserving delta function $(2\pi)^3 \delta^{(3)} (\mathbf k +\mathbf k')$ is stripped from the two point function.

There are two contributions, the first is
\begin{align}\label{firstcontribution}
\langle \zeta_{\mathbf k} \zeta_{-\mathbf k} \rangle'^{(1)} = 2 |c_2|^2 \frac{e^{i\pi\kappa}}{32 k^3 M_{\rm pl}^4\epsilon^2} \bigg| \int_{0}^{\infty} dx_1  \frac{e^{i x_1} W_{\kappa,\mu}(-2ix_1)}{x_1}  \bigg|^2 + (\kappa\rightarrow - \kappa)~.
\end{align}
The indefinite integral can be integrated directly, which yields
\begin{align}\label{meijer}
\mathcal I =\int d x \frac{e^{i x} W_{\kappa,\mu}(-2ix)}{x} = G_{2,3}^{2,1}\left(-2 i x \left|
\begin{array}{c}
1,-\kappa +1 \\
\frac{1}{2}-\mu ,\mu +\frac{1}{2},0 \\
\end{array}
\right.\right)~,
\end{align}
where $G$ is the Meijer function defined through the Gamma function in the following way
\begin{align}
	G_{p,q}^{m,n}\left(x \left|
	\begin{array}{c}
	a_1,\ldots,a_p \\
	b_1,\ldots,b_q \\
	\end{array}
	\right.\right) \equiv \frac{1}{2\pi i} \int_{\gamma_L} \frac{\prod_{j=1}^m \Gamma(b_j-s) \prod^n_{j=1} \Gamma(1-a_j+s) }{\prod_{j=n+1}^p \Gamma(a_j-s) \prod^q_{j=m+1} \Gamma(1-b_j+s) } x^s d s~.
\end{align}
At $x=0$, the integral $\mathcal I$ gives $0$. At $x\rightarrow\infty$, it gives 
\begin{align}\label{nontimeorderepower}
	\frac{\Gamma(\frac{1}{2}-\mu)\Gamma(\frac{1}{2}+\mu)}{\Gamma(1-\kappa)}~.
\end{align}
The second contributions is
\begin{align}\label{timeorderedintegralpower}
\langle \zeta_{\mathbf k} \zeta_{-\mathbf k} \rangle'^{(2)} = -4  |c_2|^2 \frac{e^{i\pi\kappa}}{32 k^3 M_{\rm pl}^4\epsilon^2}  {\rm Re} \bigg[ \int_{0}^{\infty} dx_2   \frac{e^{-i x_2} W_{-\kappa,-\mu}(2ix_2)}{x_2} \int_{0}^{x_2} dx_1  \frac{e^{-i x_1} W_{\kappa,\mu}(-2ix_1)}{x_1}  \bigg]  + (\kappa\rightarrow - \kappa)~.
\end{align}
This integral is difficult to evaluate, thus, {we use series expansion to calculate the analytical expression for the integral. The detailed calculation of the power spectrum is presented in Appendix~\ref{analyps}} 
\begin{align}
\langle \zeta_{\mathbf k} \zeta_{-\mathbf k} \rangle'^{(2)} = -4  |c_2|^2 \frac{e^{i\pi\kappa}}{32 k^3 M_{\rm pl}^4\epsilon^2}  {\rm Re}[ \mathcal{I}_2]  + (\kappa\rightarrow - \kappa)~,
\end{align}
{where $\mathcal{I}_2 = P_1 + P_2 + P_3 + P_4$. We present the expression of each component explicitly}
\begin{align}
P_1 &= \frac{\Gamma (2 \mu)\Gamma (-2 \mu)}{\Gamma (\frac{1}{2} + \mu + \kappa)\Gamma (\frac{1}{2} - \mu - \kappa)} \sum_{n=0}^{\infty}\sum_{m=0}^{\infty}\frac{(-1)^{\frac{1}{2}+ \mu+ n + 2 m} 2}{n ! m! (\frac{1}{2} + \mu +m)} \Gamma (1 +m +n ) \\\nonumber
&\,_2F_1\left(- n, -\mu + \kappa + \frac{1}{2}; 1 - 2 \mu ; 2\right) \,_2F_1\left(- m, \mu - \kappa + \frac{1}{2}; 2 \mu + 1; 2\right)\,_2F_1 \left(1 +m +n ,\frac{1}{2} + \mu +m ; \frac{3}{2} + \mu +m ;-1\right)~,\\
P_2 &= \frac{\Gamma (2 \mu)\Gamma(2 \mu)}{\Gamma (\frac{1}{2} + \mu + \kappa)\Gamma (\mu - \kappa + \frac{1}{2})} \sum_{n=0}^{\infty}\sum_{m=0}^{\infty}\frac{(-1)^{\frac{1}{2} - \mu + n + 2 m} 2^{1- 2\mu}}{n ! m ! (\frac{1}{2} - \mu +m)} \Gamma (1 - 2 \mu +m +n)\\\nonumber
& \,_2F_1\left(- n, -\mu + \kappa + \frac{1}{2}; 1 - 2 \mu ; 2\right) \, _2F_1 \left(-m, -\mu - \kappa+ \frac{1}{2};2 \mu + 1;2\right)\,_2F_1 \left(1 - 2 \mu +m +n,\frac{1}{2} - \mu +m; \frac{3}{2} - \mu +m;-1\right)~,\\
P_3 &= \frac{\Gamma(-2 \mu)\Gamma (-2 \mu)}{\Gamma (-\mu + \kappa + \frac{1}{2})\Gamma (\frac{1}{2} - \mu - \kappa)}\sum_{n=0}^{\infty} \sum_{m=0}^{\infty}\frac{(-1)^{\frac{1}{2}+\mu + n + 2 m} 2^{1 + 2 \mu}}{n ! m ! (\frac{1}{2} + \mu +m)} \Gamma (1 + 2\mu +m +n)\\\nonumber
&\, _2F_1 \left(-n, \mu + \kappa+ \frac{1}{2};1 - 2 \mu;2\right) \,_2F_1\left(- m, \mu - \kappa + \frac{1}{2}; 2 \mu + 1; 2\right)\,_2F_1 \left(1 + 2\mu +m +n,\frac{1}{2} + \mu +m; \frac{3}{2} + \mu +m;-1\right)~,\\
P_4 &= \frac{\Gamma(-2 \mu)\Gamma(2 \mu)}{\Gamma (-\mu + \kappa + \frac{1}{2})\Gamma (\mu - \kappa + \frac{1}{2})}\sum_{n=0}^{\infty} \sum_{m=0}^{\infty}\frac{(-1)^{\frac{1}{2} -\mu+ n + 2 m} 2}{n ! m ! (\frac{1}{2} - \mu +m)} \Gamma (1 +m +n)\\\nonumber
& \, _2F_1 \left(-n, \mu + \kappa+ \frac{1}{2};1 - 2 \mu;2\right) \, _2F_1 \left(-m, -\mu - \kappa+ \frac{1}{2};2 \mu + 1;2\right) \,_2F_1 \left(1 +m +n,\frac{1}{2} - \mu +m; \frac{3}{2} - \mu +m;-1\right)~.\nonumber
\end{align}

{We can see that $m = 0 , n = 0$ of $P_1$ contributes to the leading order term of $\mathcal{I}_2$   in FIG.\ref{leading}. We then would apply this approximation in calculating $\mathcal{I}_2$,
}
\begin{align}
\mathcal{I}_2 &\approx  \frac{\Gamma (2 \mu)\Gamma (-2 \mu)}{\Gamma (\frac{1}{2} + \mu + \kappa)\Gamma (\frac{1}{2} - \mu - \kappa)} \frac{(-1)^{\frac{1}{2}+ \mu} 2}{ (\frac{1}{2} + \mu )} \\\nonumber
&\,_2F_1\left(0 , -\mu + \kappa + \frac{1}{2}; 1 - 2 \mu ; 2\right) \,_2F_1\left( 0 , \mu - \kappa + \frac{1}{2}; 2 \mu + 1; 2\right)\,_2F_1 \left(1 ,\frac{1}{2} + \mu ; \frac{3}{2} + \mu ;-1\right) \\\nonumber
& = \frac{i e^{-i \pi  \mu }}{2 \mu } \csc (2 \pi  \mu ) \cos (\pi  (\kappa +\mu )) \left(\psi ^{(0)}\left(\frac{\mu}{2}+\frac{1}{4}\right)-\psi ^{(0)}\left(\frac{\mu }{2}+\frac{3}{4}\right)\right)~.
\end{align}

\begin{figure}[htbp] 
	\centering 
	\includegraphics[width=12cm]{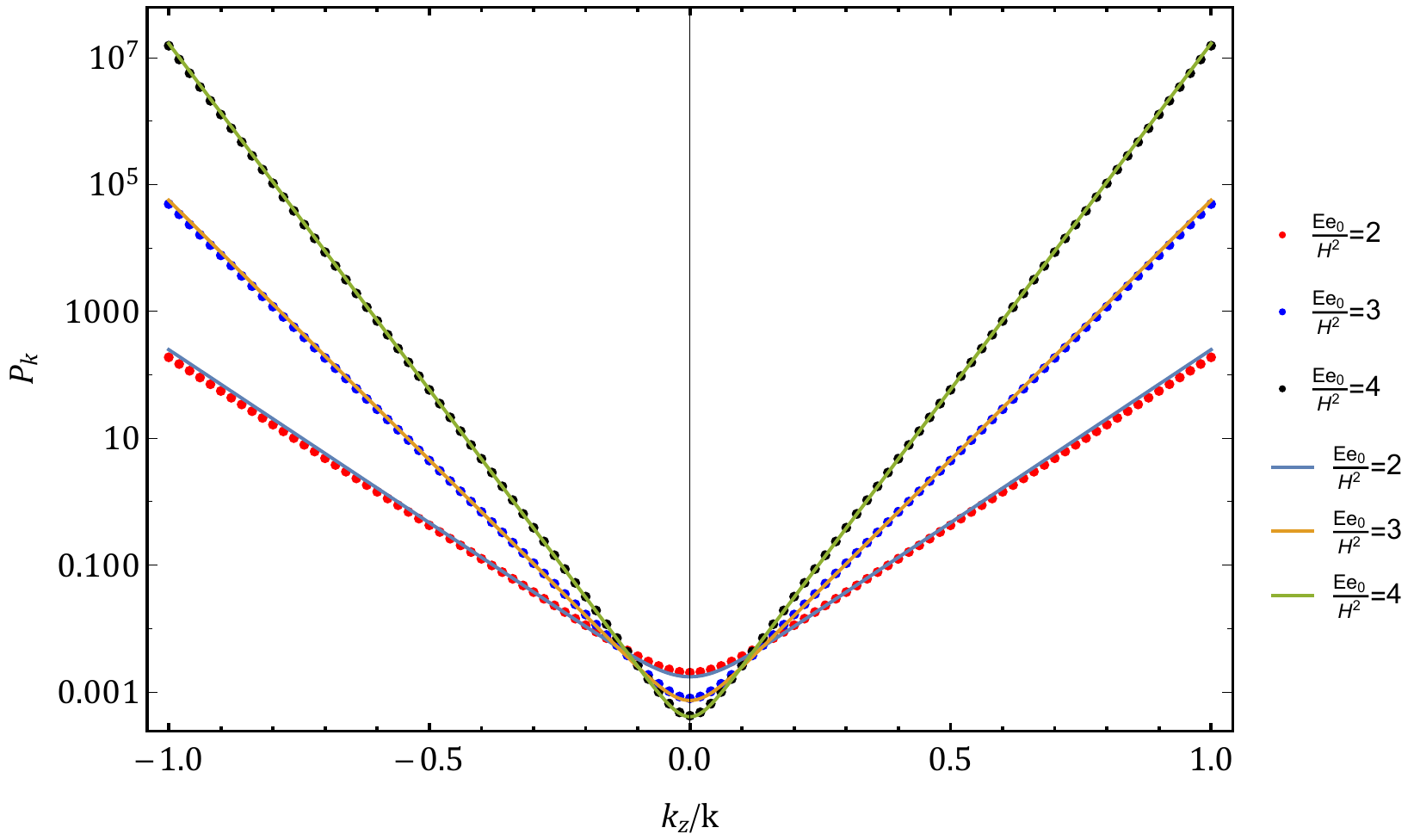}   
	\caption{{We here make a comparison on the analytical (line plot) and numerical (dotted plot)  results for different electric field strength. } At different electric field strength, the power spectrum has different angular dependence. We set the mass of massive field to be $3H/2$. When the orientation is aligned to the orientation of the field, the power spectrum is much larger. If the orientation is perpendicular to the orientation of the field, although the electric field is strong, Schwinger effect is weak and the effective mass is large enough to suppress the power spectrum, hence, the power spectrum can be small with strong electric field strength. At the orientation parallel to the field's orientation, the Schwinger effect is strong enough to get a large power spectrum.} \label{angular}
\end{figure}

The power spectrum $P_\zeta$ is obtained as
\begin{align}
\langle \zeta_{\mathbf k} \zeta_{-\mathbf k} \rangle' = \langle \zeta_{\mathbf k} \zeta_{-\mathbf k} \rangle'^{(1)} + \langle \zeta_{\mathbf k} \zeta_{-\mathbf k} \rangle'^{(2)} \equiv  \frac{2\pi^2}{k^3} P_\zeta (\mathbf{k})~.
\end{align}  
From here and the following, when making the plot, we set $c_2=c_3=M_{\rm pl}=\epsilon=H=1$. We plot the angular dependence of the power spectrum in FIG.~\ref{angular}. The produced charged particles can leave non trivial angular dependence on the power spectrum. The power spectrum grows exponentially as the quantity $k_z/k$ increases. This signature is understood as the production of virtual particles increases exponentially when the momentum of the positive charged  massive scalar particle is aligned with the electric field $E$ whereas the momentum of the negative charged massive scalar particle is opposite to the direction of the electric field $E$. This signature is a unique signature which cannot be generated by other mechanisms to the knowledge we know of. 

\begin{figure}[htbp] 
	\centering 
	\includegraphics[width=12cm]{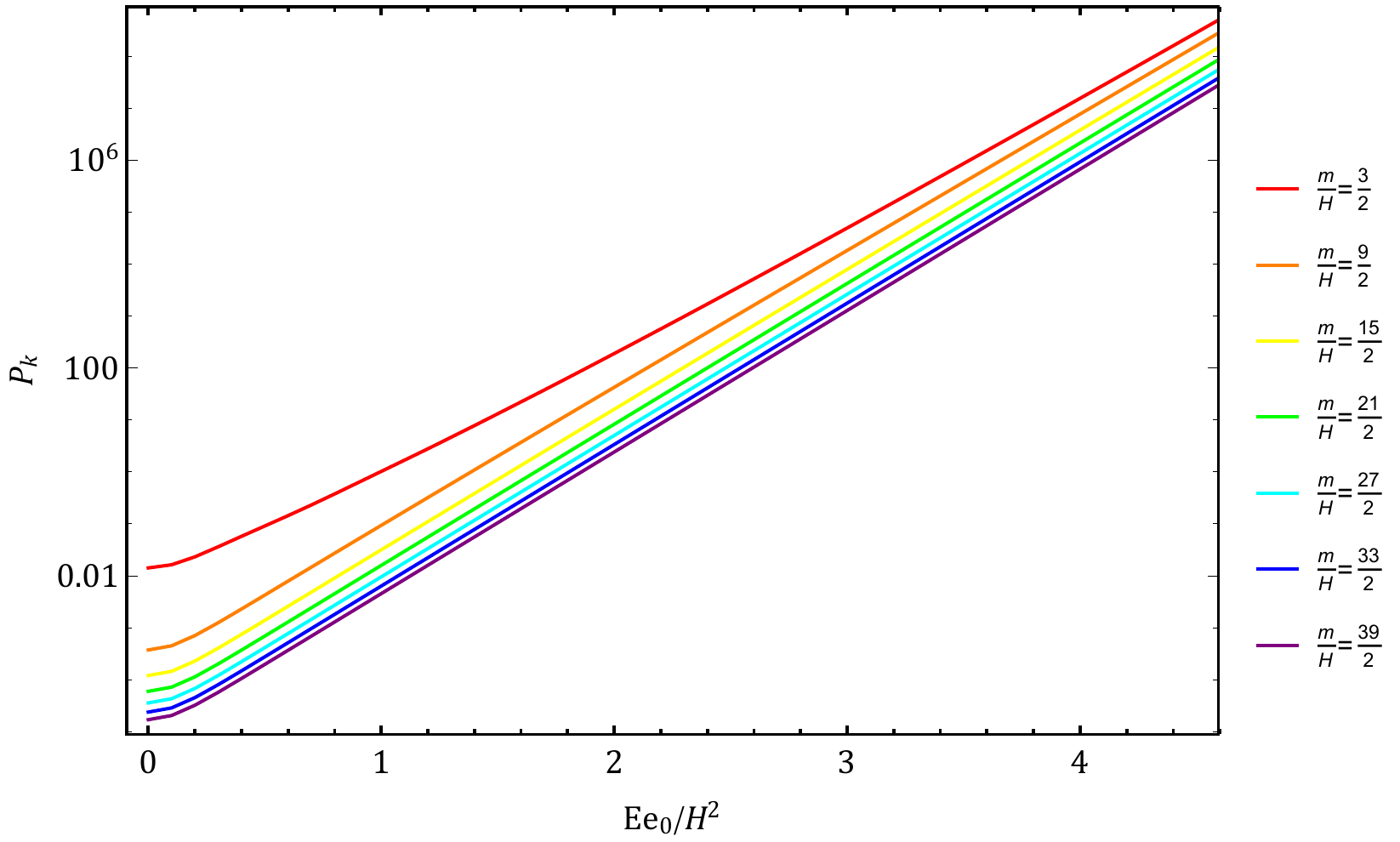}   
	\caption{We set $k_{z}/k=1$ and the mass of the massive field to be $3H/2, 9H/2, 15H/2, 21H/2, 27H/2, 33H/2, 39H/2$ respectively. The power spectrum increases exponentially with respect to the electric field strength.} \label{edependence}
\end{figure}

\begin{figure}[htbp] 
	\centering 
	\includegraphics[width=12cm]{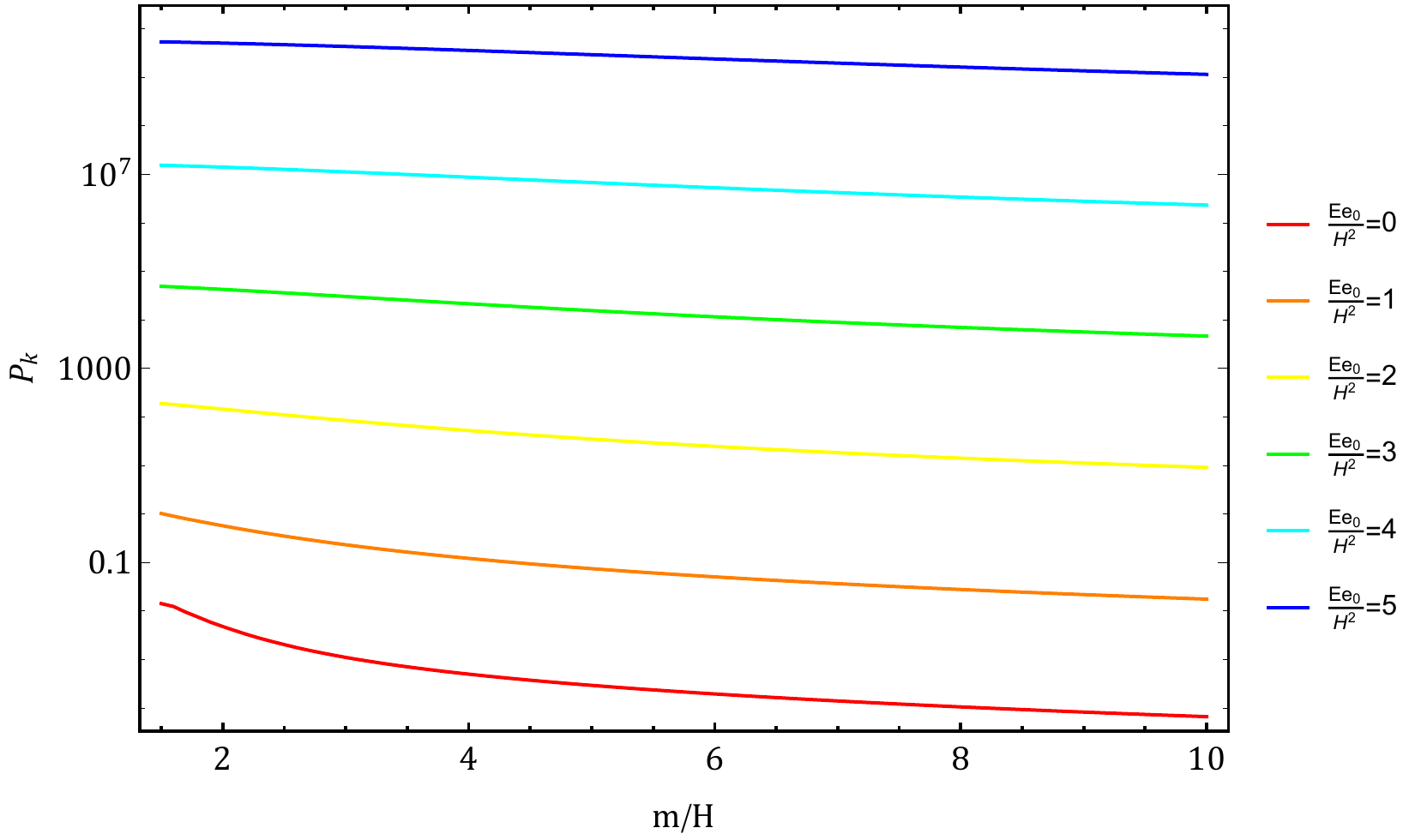}   
	\caption{We set $k_{z}/k=1$ and $\kappa=0, -i, -2i, -3i, -4i, -5i, -6i$ respectively. For small electric field strength, the power spectrum decreases following the power $1/|\mu|^2$, whereas when the electric field strength increases, the power spectrum decays exponentially when the mass increases.} \label{mdependence}
\end{figure}

We also plot the dependence of the power spectrum on electric field strength in FIG.~\ref{edependence} and the dependence of the power spectrum on the mass of massive field in FIG.~\ref{mdependence}. 

\section{Bispectrum}\label{Bispectrum}
\begin{figure}[htbp] 
	\centering 
	\includegraphics[width=4.5cm]{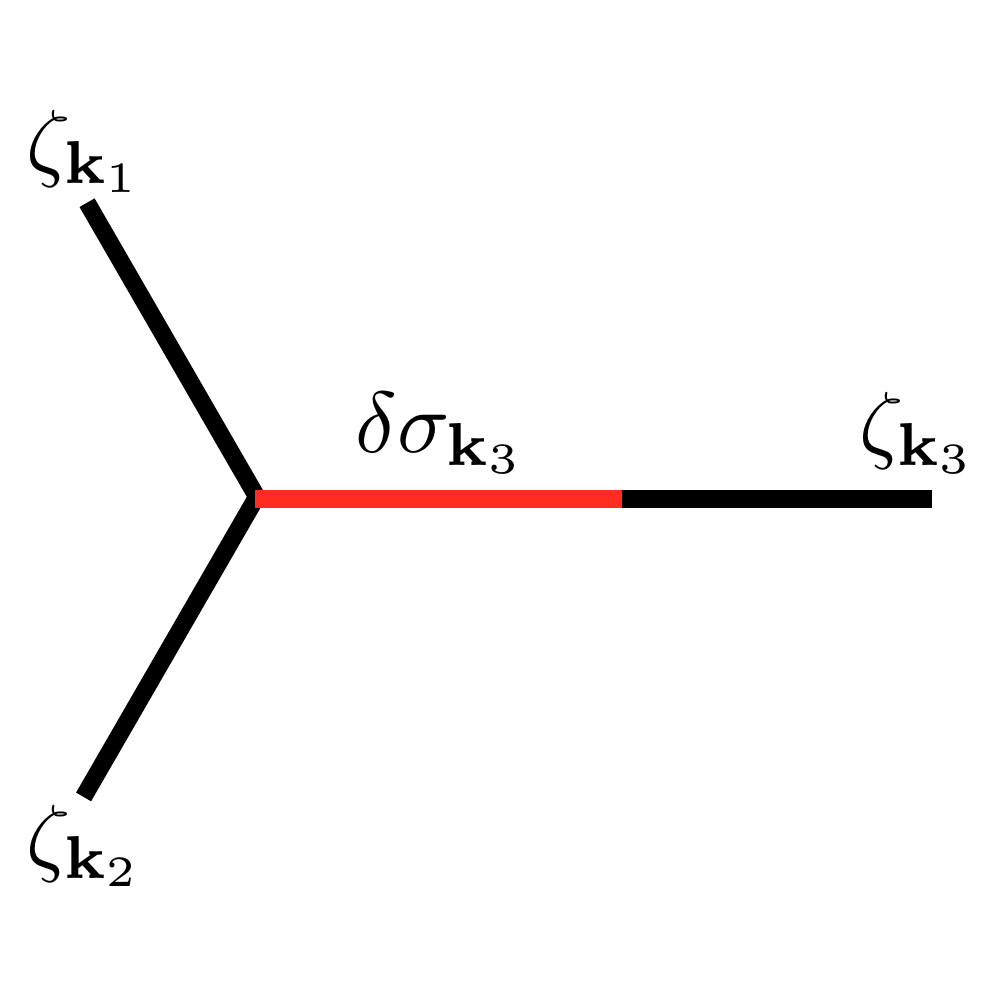}   
	\caption{{The Feynman diagram of the bispectrum that involves the interaction of $\delta\sigma$ with the inflaton. For the total contribution, we need to include the correction from $\delta\sigma^*$.}} \label{fenman1}
\end{figure} 
In this section, we study the bispectrum of this model, {which is shown in FIG. \ref{fenman1}}. For the bispectrum, using the Schwinger-Keldysh formalism \cite{Chen:2017ryl}, the bispectrum can be expressed as
\begin{align}
\langle  \zeta_{\mathbf k_1} \zeta_{\mathbf k_2}\zeta_{\mathbf k_3} \rangle'  &=  c_2c^*_3 \sum_{a,b = \pm} ab \int_{-\infty}^{0}   \int_{-\infty}^{0}   \frac{d\tau_1}{(-H\tau_1)^3} \frac{d\tau_2}{(-H\tau_2)^2} \partial_{\tau_1}G_{a+}({\mathbf k_3},\tau_1,0) D_{ab}({\mathbf k _3},\tau_1,\tau_2) \partial_{\tau_2}G_{b+}({\mathbf k _1},\tau_2,0)\partial_{\tau_2}G_{b+}({\mathbf k_2},\tau_2,0) \nonumber\\
&+ ({\kappa\rightarrow-\kappa},c_2 \rightarrow c_2^*,c_3^* \rightarrow c_3 )~.
\end{align}
There are three contributions
\begin{align}\label{bispectrum3}
\langle \zeta_{\mathbf k_1} \zeta_{\mathbf k_2}\zeta_{\mathbf k_3} \rangle' = \langle \zeta_{\mathbf k_1} \zeta_{\mathbf k_2}\zeta_{\mathbf k_3} \rangle'^{(1)} + \langle \zeta_{\mathbf k_1} \zeta_{\mathbf k_2}\zeta_{\mathbf k_3} \rangle'^{(2)} + \langle \zeta_{\mathbf k_1} \zeta_{\mathbf k_2}\zeta_{\mathbf k_3} \rangle'^{(3)} +{\rm 5\,\, Permutations}+ ({\kappa\rightarrow-\kappa},c_2 \rightarrow c_2^*,c_3^* \rightarrow c_3 )~,
\end{align}
where $\langle \zeta_{\mathbf k_1} \zeta_{\mathbf k_2}\zeta_{\mathbf k_3} \rangle'^{(1)} $, $\langle \zeta_{\mathbf k_1} \zeta_{\mathbf k_2}\zeta_{\mathbf k_3} \rangle'^{(2)} $ and $\langle \zeta_{\mathbf k_1} \zeta_{\mathbf k_2}\zeta_{\mathbf k_3} \rangle'^{(3)} $ are computed as
\begin{align}\nonumber
&\langle \zeta_{\mathbf k_1} \zeta_{\mathbf k_2}\zeta_{\mathbf k_3} \rangle'^{(1)} =- 2 c_2c^*_3 \frac{H^3 e^{i\pi\kappa}}{128 \epsilon^3 k_1 k_2 k^4_3 M_{\rm pl}^6}   \int_{0}^{\infty} dx_1  \frac{e^{i x_1} W_{\kappa,\mu}(-2ix_1)}{x_1} \int_{0}^{\infty} dx_2 x_2 e^{-i \frac{k_1+k_2}{k_3} x_2}  W_{-\kappa,-\mu} (2 i  x_2 )~. \\\nonumber
&\langle \zeta_{\mathbf k_1} \zeta_{\mathbf k_2}\zeta_{\mathbf k_3}\rangle'^{(2)} = 2  c_2c^*_3 \frac{H^3 e^{i\pi\kappa}}{128 \epsilon^3 k_1 k_2 k^4_3 M_{\rm pl}^6}   {\rm Re} \bigg[ \int_{0}^{\infty} dx_2 x_2  e^{-i  \frac{k_1+k_2}{k_3} x_2} W_{-\kappa,-\mu}(2i x_2) \int_{0}^{x_2} dx_1  \frac{e^{-i x_1} W_{\kappa,\mu}(-2ix_1)}{x_1}  \bigg] ~. \\\nonumber
&\langle \zeta_{\mathbf k_1} \zeta_{\mathbf k_2}\zeta_{\mathbf k_3}\rangle'^{(3)} = 2  c_2c^*_3 \frac{H^3 e^{i\pi\kappa}}{128 \epsilon^3 k_1 k_2 k^4_3 M_{\rm pl}^6}   {\rm Re} \bigg[ \int_{0}^{\infty} dx_2 \frac{e^{-i x_2} W_{-\kappa,-\mu}(2ix_2)}{x_2}  \int_{0}^{x_2} dx_1  x_1  e^{-i  \frac{k_1+k_2}{k_3} x_1} W_{\kappa,\mu}(-2i x_1) \bigg]~.
\end{align}
\begin{figure}[htbp] 
	\centering 
	\includegraphics[width=6cm]{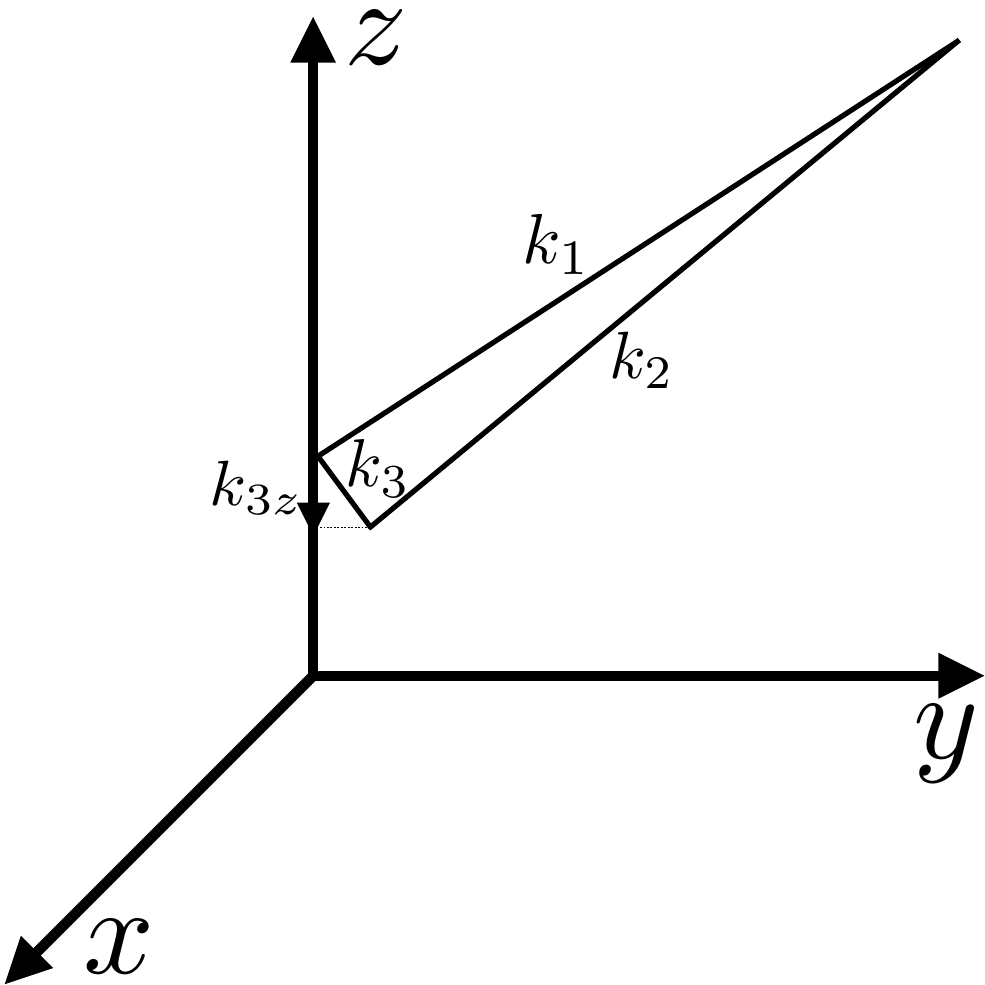}   
	\caption{This figure shows how we measure the non-Gaussianity. The $z$ direction denotes the direction of the electric field. When we measure the non-Gaussianity, we should fix the ratio of the long wavelength momentum and the short wavelength momentum $k_1/k_3$. In the meanwhile, we measure the angular dependence of the non-Gaussianity for different $k_{3z}/k_3$. $k_{3z}$ is the magnitude of the long wavelength momentum projected onto the $z$ direction. } \label{measure}
\end{figure}
\begin{figure}[htbp] 
	\centering 
	\includegraphics[width=12cm]{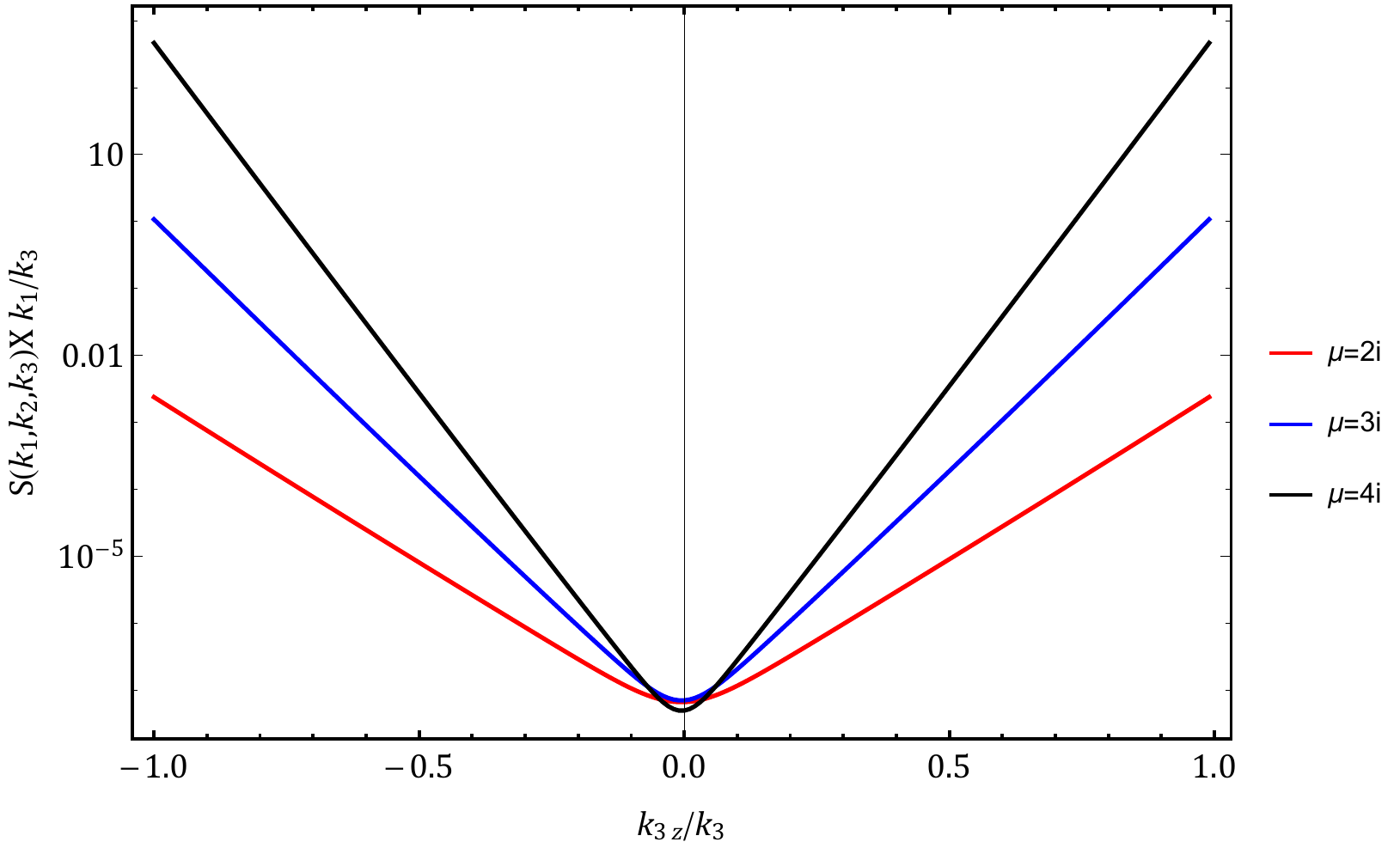}   
	\caption{This figure shows the amplified shape function $S(k_1,k_2,k_3)\times k_1/k_3$ as a function of the angular dependence of the soft momentum $k_{3z}/k_3$. In this figure, we set $m=3H/2$ and $k_1/k_3=100$. When the orientation is aligned to the orientation of the field, the amplitude of the bispectrum is much larger. At the orientation perpendicular to the field's orientation, although the electric field strength is strong, Schwinger effect is weak and the effective mass is large enough to suppress the bispectrum. At the orientation parallel to the orientation of the field, the Schwinger effect is strong enough to generate a bispectrum of large amplitude. } \label{bispectrumss}
\end{figure}
We can define the bispectrum in the form of dimensionless shape function $S(k_{1},k_{2},k_{3})$ \cite{Chen:2018xck}, defined as,
\begin{align} \label{shapefunction}
\langle \zeta_{\mathbf k_1} \zeta_{\mathbf k_2}  \zeta_{\mathbf k_3}  \rangle'   \equiv (2 \pi)^4 S(k_{1},k_{2},k_{3}) \frac{1}{(k_{1}k_{2}k_{3})^2} P^{(0)2}_{\zeta}~,
\end{align}
where $P^{(0)}_\zeta = \frac{1}{8\pi^2 {M_{\rm pl}^2}} \frac{H^2}{\epsilon}$ is the power spectrum for the curvature perturbation without the correction coming from massive fields. 

The bispectrum can be evaluated using numerical integration. We plot the angular dependence of the amplified bispectrum shape function $(k_1/k_3)\times S(k_1,k_2,k_3)$ as a function of $k_{3z}/k_3$ in FIG.~\ref{bispectrumss}. The bispectrum increases exponentially with increasing absolute value of $k_{3z}/k_3$. When $k_{3z}/k_3$ is positive, the main contribution comes from the positive charged particles whereas when $k_{3z}/k_3$ is negative, the main contribution comes from the negative charged particles.

\begin{figure}[htbp] 
	\centering 
	\includegraphics[width=10cm]{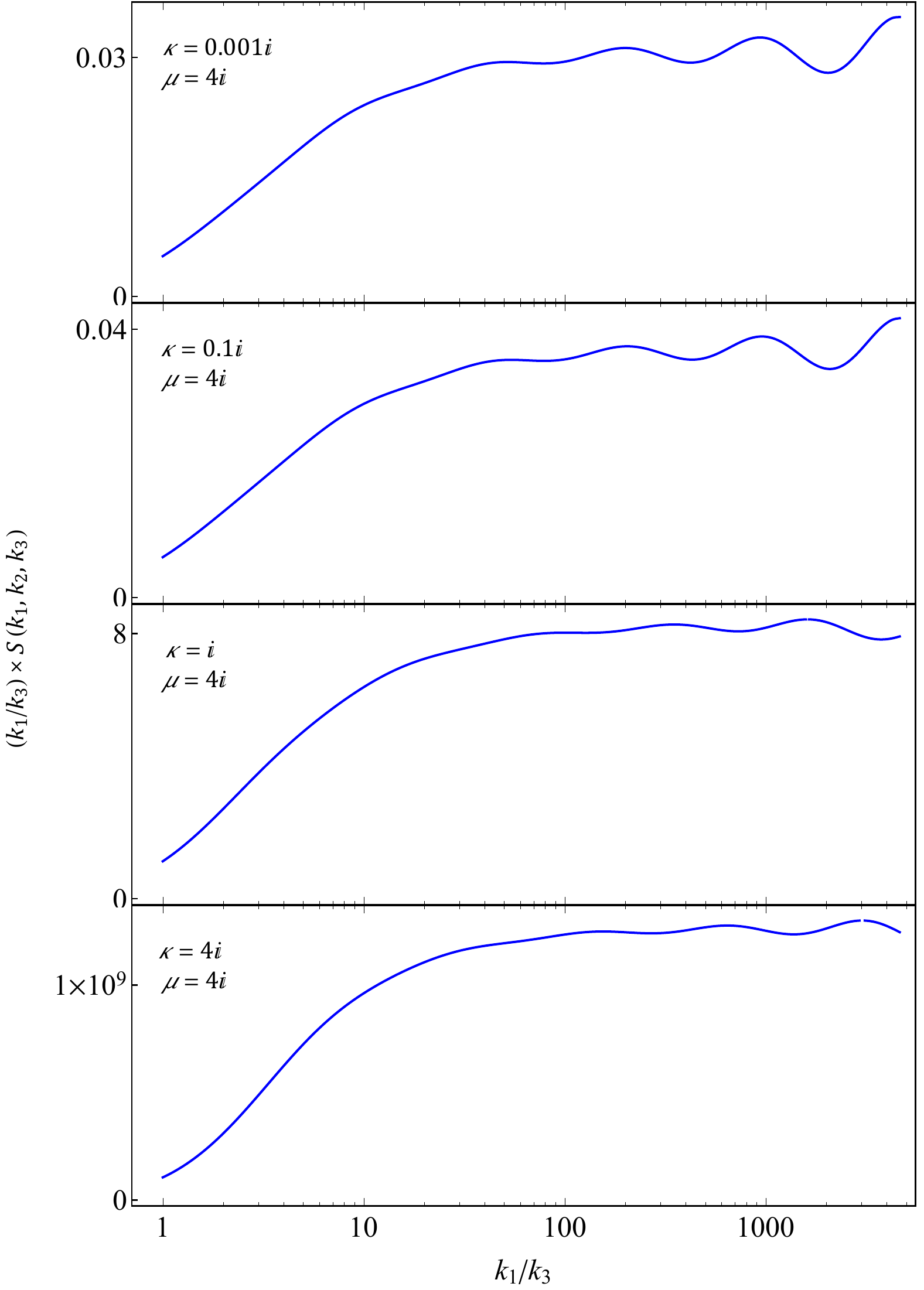}   
	\caption{The figure shows the clock signal from the bispectrum. Here we set the momentum to be in the z-direction and hence we set $k_{3z}/k_3=1$. We can see the suppression of the clock signal as the electric field increases. The frequency of the clock signal remains unchanged as the change in velocity of the produced particles occurs in the subhorizon and remains stationary during horizon crossing. Hence, no change in frequency of the oscillatory signal would be seen. However, the amplitude of the oscillatory signal would increase as the magnitude of the electric field strength increase. } \label{bispectrumclock}
\end{figure}

\begin{figure}[htbp] 
	\centering 
	\includegraphics[width=10cm]{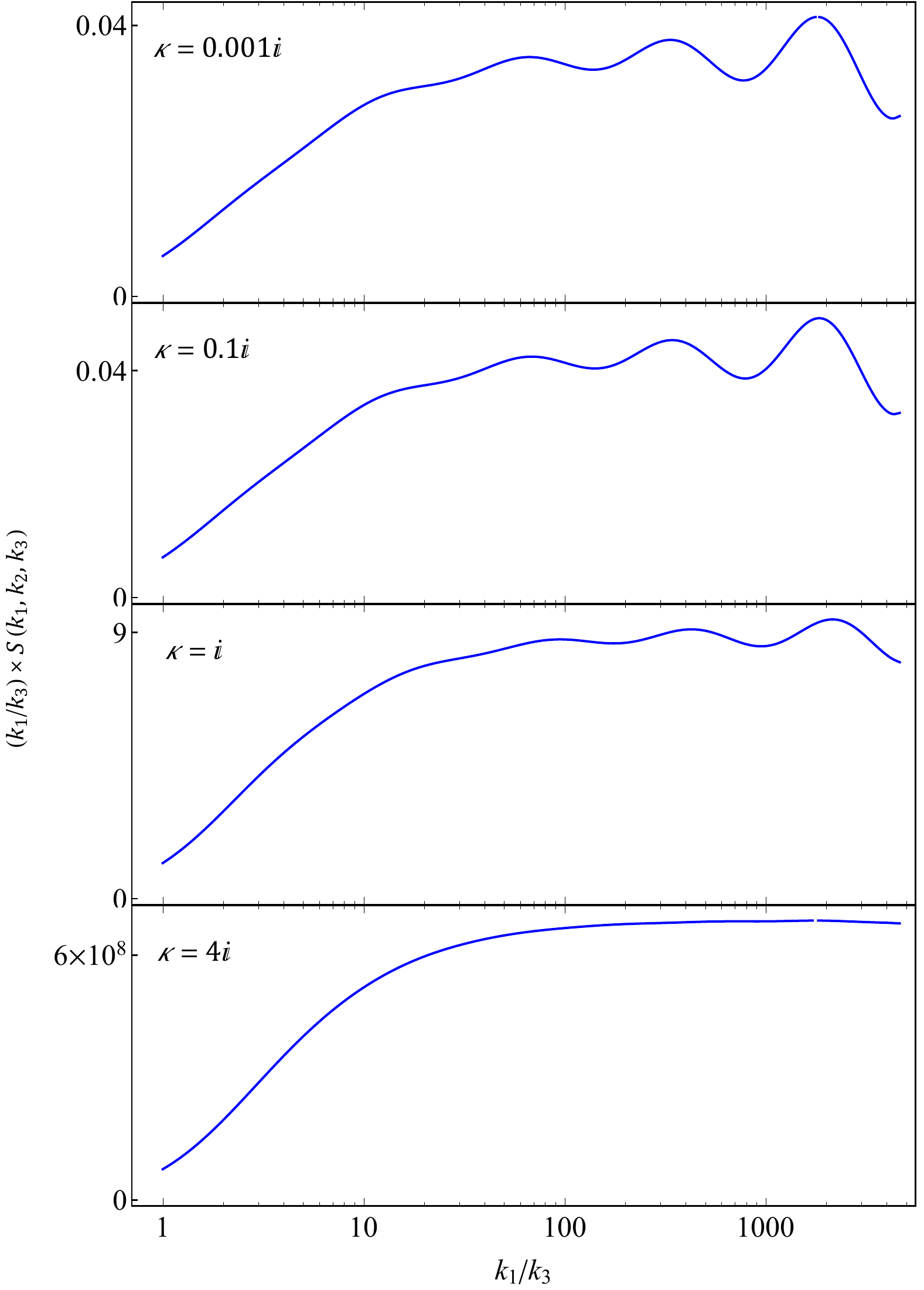}   
	\caption{The figure shows the clock signal from the bispectrum for $m=4H^2$. Here we set $k_{3z}/k_3=1$ again. We can see the suppression of the clock signal as the electric field increases. } \label{bispectrumclockconstm}
\end{figure}

We plot the clock signals of the squeeze limit of the bispectrum with fixed effective mass $\mu$ in FIG.~\ref{bispectrumclock} and with fixed mass $m$ in FIG.~\ref{bispectrumclockconstm}. In both cases, we see that the clock signal is less obvious when the strength of the electric field increases. In the case of fixing the  effective mass $\mu$ case, the absolute value of the clock signal increases with increasing electric field strength due to the enhanced particle production rate. However, the relative amplitude between the clock signal and the contribution of the non-oscillating part coming from some local process decreases. This is because the contribution of the non-oscillating part increases faster than the clock signal when the electric field strength increases. For the fixed mass $m$ case, when the electric field strength increases, the amplitude of the clock signal relative to the non-oscillating part decreases more dramatically compared with the fixed effective mass case. This is because the electric field strength contributes to the  effective mass of the charged massive scalar particles. The energy needed to produce a particle increases accordingly. The combination of these two effect causes the amplitude of the clock signal relative to the non-oscillating part barely observable starting from $\kappa = 4i$. 
The analytical expression of the clock signal in the large mass and small electric field strength can be obtained in Appendix $\ref{largemassbispectrum}$.  We are particularly interested in the squeezed limit where $k_1\sim k_2\gg k_3$. In this limit, the shape function is
\begin{align}
	S(k_1,k_3) = \frac{1}{H M_{\rm pl}^2 \epsilon} {\rm Re} \bigg[2^{\mu-9/2}c_2c_3^* f(\mu,\kappa) 
	 \bigg(\frac{k_1}{k_{3}}\bigg)^{-1/2-\mu} \bigg]+ (\kappa \rightarrow -\kappa,c_2\rightarrow c_2^*,c_3^*\rightarrow c_3 )~,
\end{align}
with the prefactor given by 
\begin{align}
f(\mu,\kappa)\equiv   e^{i \pi \kappa }  \frac{\Gamma(-2\mu)^2\Gamma\left(\frac{1}{2}+\mu\right)\Gamma\left(\frac{5}{2}+\mu\right)}{\Gamma(\frac{1}{2}-\mu-\kappa)\Gamma(\frac{1}{2}+\mu-\kappa)} (1+\text{sin} \left(\pi\mu\right))~, 
\end{align}
where the full derivation can be obtained in Appendix $\ref{clock}$. We can see that in the large mass limit, all the $\Gamma$ functions contribute a factor of $e^{-2 \pi |\mu|}$ and sin($\pi\mu$) would give a contribution of $e^{ \pi |\mu|}$. Hence, the Boltzmann suppression factor $e^{-\pi |\mu|}$ is recovered in this limit. At the end of this section, we would like to compare several mechanisms that can generate large clock signals even if the mass of the $\sigma$ field are large. There are a few categories of mechanisms listed as the following.
\begin{itemize}
	\item The presence of a new scale. In \cite{Flauger:2016idt}, non-adiabatic production of very heavy fields is studied. The signatures of this model can be large due to the existence of another scale $\dot\phi$ with $\phi$ as the inflaton.  
	\item Finite temperature effect. In \cite{Tong:2018tqf}, the clock signal of the quasi-single field inflation is studied in the context of warm inflation. The particle production rate can be unsuppressed when the effective mass of the particle is changed due to the finite temperature effect. 
	\item The presence of chemical potential. In \cite{Adshead:2015kza,Adshead:2018oaa,Chen:2018xck}, the effect of the chemical potential is studied. The chemical potential can assist the production of the massive particles during inflation thus leaving a less suppressed clock signal. The mechanism we studied here also belongs to this category. However, our studies shows that although it is promising to generate a larger clock signal, one may worry that the contribution from the non-oscillating part will also increase. 
	\item Non-trivial sound speed. The non-trivial sound speed of the massive field is studied in \cite{Chen:2018sce}. The magnitude of the clock signal can also be larger than expected when the ratio of sound speed of the massive field and the inflaton is less than one. In \cite{Noumi:2012vr,Lee:2016vti}, the non-trivial sound speed of the inflaton is studied. It is shown in \cite{Lee:2016vti} that when the sound speed of the inflaton is close to zero, there will also be a change in the suppression factor of the clock signal. 
\end{itemize}

\section{Loop Correction to Bispectrum}\label{loopBispectrum}
In this section, we investigate loop corrections coming from the extra massive fields to the primordial non-Gaussianities. The technique of dealing loop correction in quasi-single field inflation can be found in  \cite{Arkani-Hamed:2015bza,Chen:2016nrs,Chen:2016uwp,Chen:2016hrz,Wu:2017lnh}. The non-oscillatory part of the diagram is usually UV divergent and we need a systematic way of regularization and renormalization following \cite{Weinberg:2005vy,Senatore:2009cf,Senatore:2012nq,Pimentel:2012tw,Dimastrogiovanni:2018uqy}. Luckily, the clock signal is free from UV divergence and we can evaluate it easily.

\begin{figure}[htbp] 
	\centering 
	\includegraphics[width=4.5cm]{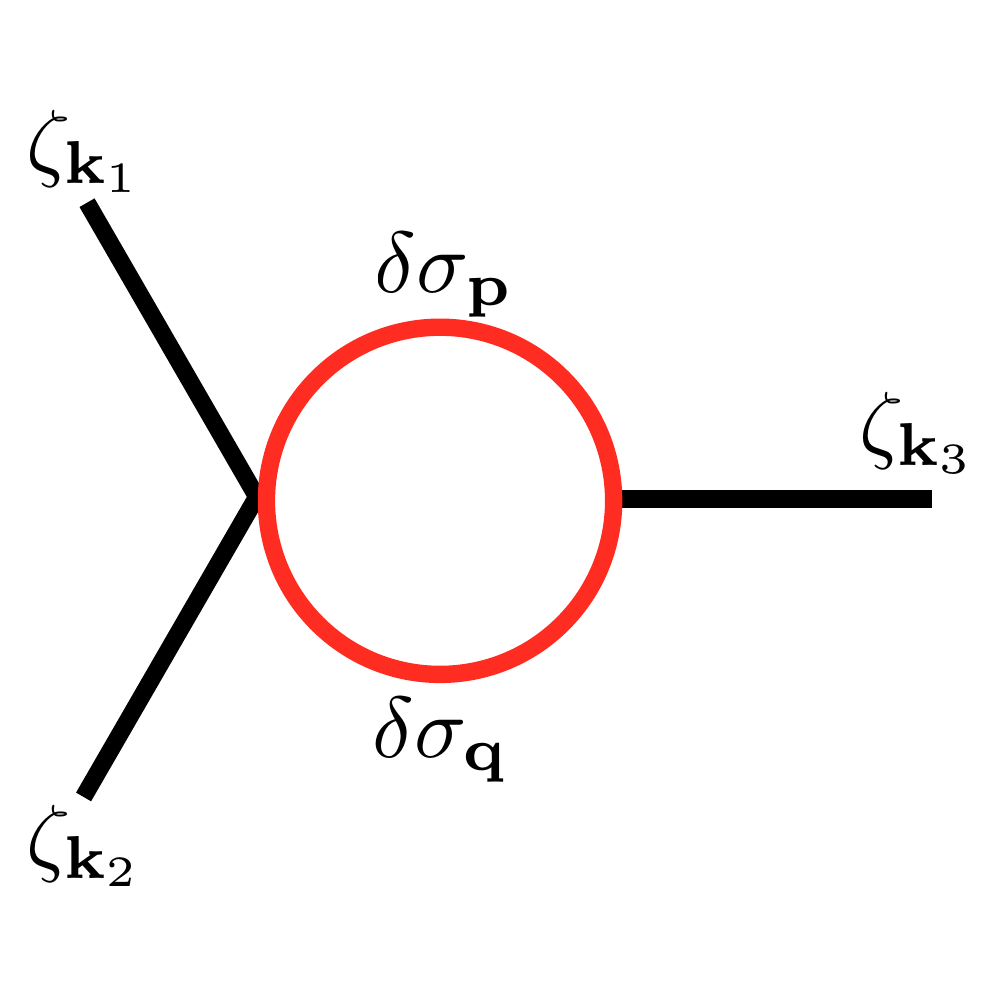}   
	\caption{The Feynman diagram we are considering {for the loop correction to the bispectrum.}} \label{fenman}
\end{figure}

Using the Schwinger-Keldysh formalism, the bispectrum corresponding FIG.~\ref{fenman} can be obtained as
\begin{align}\nonumber
\langle  \zeta_{\mathbf k_1} \zeta_{\mathbf k_2}\zeta_{\mathbf k_3} \rangle'  &=  c^{\prime}_2c^{\prime}_3 \sum_{a,b = \pm} ab \int_{-\infty}^{0}   \int_{-\infty}^{0}   \frac{d\tau_1}{(-H\tau_1)^3} \frac{d\tau_2}{(-H\tau_2)^2} \partial_{\tau_1}G_{a+}({\mathbf k_3},\tau_1,0)  \int \frac{d^3 \mathbf q}{(2\pi)^3} D_{ab}(\mathbf p,\tau_1,\tau_2) D_{ba}(\mathbf q,\tau_2,\tau_1) \\
&\quad  \partial_{\tau_2}G_{b+}({\mathbf k_1},\tau_2,0)\partial_{\tau_2}G_{b+}({\mathbf k_2},\tau_2,0) + ({\kappa\rightarrow-\kappa})~,
\end{align}
where $\mathbf p$ and $\mathbf q$ are the loop momentum that satisfies the constraint $\mathbf p+\mathbf q = \mathbf k_3$. After evaluation, we get 
\begin{align}\nonumber
\langle \zeta_{\mathbf k_1} \zeta_{\mathbf k_2}  \zeta_{\mathbf k_3}  \rangle' =   c^{\prime}_2 c^{\prime}_3 {\rm Re} \bigg[g(\mu,\kappa)\frac{ 2^{-1-8\mu}H^5   }{ k_1 k_2   k^4_{12} M_{\rm pl}^6 \epsilon^3}  \bigg(\frac{k_{12}}{ k_3}\bigg)^{2\mu} \bigg]  + ({\kappa\rightarrow-\kappa})~,
\end{align}
with the prefactor given by 
\begin{align}
g(\mu,\kappa) = e^{2 i \pi \kappa}\frac{\Gamma (2-2 \mu )\Gamma (4-2 \mu )\Gamma(-2\mu)^4}{\Gamma(1/2-\mu-\kappa)^2\Gamma(1/2+\mu-\kappa)^2}(\text{sin}(\pi\mu))^2~,
\end{align}
 and with the definition of the shape function $S(k_1,k_2,k_3)$ in \eqref{shapefunction} and taking the limit $k_1=k_2\gg k_3$, we can obtain the expression for the shape function
\begin{align}
S(k_1,k_1,k_3) = c^{\prime}_2 c^{\prime}_3{\rm Re}\bigg[ g(\mu,\kappa)\frac{ 2^{3-8\mu}H  }{  M_{\rm pl}^2 \epsilon}   \bigg(\frac{2 k_1}{ k_3}\bigg)^{2\mu-2} \bigg]  + ({\kappa\rightarrow-\kappa})~.  
\end{align}
We can see that from the loop diagram, the massless curvature modes resonates with two pairs of massive fields and generate two sets of clock signals. The  final clock signal would be contributed from the interference of these two clock signals
and hence has a frequency doubled of the tree level diagram. The total Boltzmann suppression factor is of $e^{-2 \pi \mu}$ where all the $\Gamma$ functions contribute $e^{-4 \pi \mu}$ in total and the sin functions contribute a factor of $e^{2 \pi \mu}$. The suppression for the loop correction is the square of
the tree-level case due to the excitation of the two massive fields in the loop diagram.
 
\section{Conclusion and Outlook}\label{Conclusion}
In this work, we consider the imprints of the Schwinger effect on the primordial power spectrum and bispectrum. Both the power spectrum and bispectrum obtained an angular dependence due to the fact that the electric field can assist the production of charged massive particles by adding a chemical potential to them. This angular dependence differs from other models that can too cause an angular dependence on the primoridial power spectrum and bispectrum. As a result, the production rate aligned or opposite to the direction of the charged particles gets enhanced. On the other hand, the production rate perpendicular to the direction of the electric field is suppresed due to the contribution of the electric field strength to the effective mass of the charged scalar particles. 

There are many interesting possibilities to explore. We list a few of them and hope to address some of these possibilities in the future.
\begin{itemize}
	\item The influence of the primordial magnetic field on the power spectrum and bispectrum in the context of quasi-single field inflation. The pair production of the charged scalar field in the presence of a constant electric and magnetic field is studied in \cite{Bavarsad:2017oyv}. We would like to couple the charged particles with the inflaton and see what kind of signature these particles would imprint on the primordial power spectrum and bispectrum of the curvature perturbations. These signatures on the power spectrum and bispectrum will provide supporting evidences to the existence of the primordial magnetic field. 
	\item The signature from other fields produced by Schwinger effect during inflation. Schwinger effect not only produces charged scalar particles, but charged fermions too \cite{Hayashinaka:2016qqn}. The production rate is very similar. However, the production of fermions may lead to other types signatures on the power spectrum and bispectrum. 
	\item Other sources like $SU(2)$ gauge fields \cite{Lozanov:2018kpk,Maleknejad:2018nxz,Dimastrogiovanni:2016fuu}. In this case, the production rate is suppressed as the interaction strength increases. However, some signal of cosmological collider type can be generated by the spin-2 field which is required by this type of model. 
\end{itemize}

\section*{Acknowledgments} 
We thank Guillem Domenech, Xi Tong, Yuhang Zhu and Henry Tye for useful discussions. This work is supported in part by ECS Grant 26300316 and GRF Grant 16301917 from the Research Grants Council of Hong Kong. WZC is supported by the Targeted  Scholarship Scheme under the HKSAR Government Scholarship Fund and the Kerry Holdings Limited Scholarship.

\appendix 
\section{Backreaction due to Conformal Coupling}\label{backreaction}
{We investigate the effect of the backreaction on the background equations of motion. This effect is also studied in (1+1)D in \cite{Stahl:2016geq} and general dimension in \cite{Bavarsad:2016cxh,Sobol:2018djj}. We consider a modified quasi-single field inflation model where a charged isocurvaton $(\sigma)$ is coupled to the inflaton $(\theta)$ \cite{Chen:2009we,Chen:2009zp}. }
\begin{align}\label{action_f}
S= \int d^4 x \sqrt{-g} \bigg[ -g^{\mu\nu}(\tilde{R}+\sigma)(\tilde{R}+\sigma^*)\partial_\mu \theta \partial_\nu \theta- g^{\mu\nu} D_\mu \sigma^* D_\nu \sigma - V_{\text{sr}}(\theta) - m^2 |\sigma|^2-\frac{1}{4}f^2(\theta) F_{\mu\nu}F^{\mu\nu}  \bigg]~,
\end{align} 
where $D_\mu \equiv \partial_\mu + i e(\theta_0) A_\mu$. Here we consider the charge to be dependent on the inflaton field. $F_{\mu\nu} = \partial_{\mu} A_\nu - \partial_\nu A_{\mu} $ is the electromagnetic tensor. 
{We consider the FRW metric as stated previously. The dilatonic coupling $f(\theta)$ would prevent the decay of the gauge field energy density from the expansion of the universe.  We consider the background gauge field to have the form $(0,0,0,A_z(t))$. We now have the following Hubble and continuity equation}
\begin{align}\label{hubble}
3M_{\text{pl}}^2H^2 &=  \dot{\theta}_0^2 R^2+ m^2\sigma_0^2 +V_{\text{sr}}+\frac{1}{2}\frac{f(\theta_0)^2}{a^2}\dot{A}_z^2+\frac{e(\theta_0)^2\sigma_0^2A_z^2}{a^2}~, \nonumber\\
-M_{\text{pl}}^2\dot{H} &=   \dot{\theta}_0^2R^2+\frac{1}{2}\frac{f(\theta_0)^2}{a^2}\dot{A}_z^2~,
\end{align}
{where $R = \tilde{R}+\sigma_0$. We can see that in the absence of the charge of the complex scalar field, the gauge fields act as radiation. The equation of motion reads}
\begin{align}\label{background_eqm}
\sigma_0 &= \text{constant}, \qquad \dot{\theta}_0^2 R = m^2 \sigma_0 + \frac{e(\theta_0)^2 \sigma_0 A_z^2}{a^2}, \nonumber\\
R^2 \ddot{\theta}_0 + 3R^2H \dot{\theta}_0 + \frac{V'_{\text{sr}}}{2} &= \frac{1}{2}\frac{f(\theta_0) f'(\theta_0)\dot{A}_z^2}{a^2}-\frac{e(\theta_0)e'(\theta_0)\sigma_0^2 A_z^2}{a^2}~.
\end{align}
{We can also vary (\ref{action_f}) with respect to the gauge field and obtain the corresponding Maxwell equation,}
\begin{align}\label{maxwell}
\partial_t(f(\theta_0)^2 a \dot{A}_z) = -2 e(\theta_0)^2 \sigma_0^2 a A_z~.
\end{align}
{From (\ref{hubble}), we can define the ratio of energy density of gauge fields to the inflaton,}
\begin{align}\label{Edensity_ratio}
R_1 = \frac{\dot{A}_z^2 f(\theta)^2}{2a^2 m^2\sigma_0^2}, \quad R_2 = \frac{e(\theta_0)^2  A_z^2}{a^2 m^2 }~.
\end{align}
{Since we require inflation to last for a long enough period of time, we require the gauge field energy density to be much smaller than the total energy density. This is because inflation has to be mainly driven by the inflaton itself and we require $R_1,R_2\ll 1$ to prevent inflation from ceasing \cite{Firouzjahi:2018wlp,Emami:2015qjl}. From (\ref{maxwell}), we can see that $e(\theta_0)^2 \sigma_0^2 A^2 a^{-2}$ must be negligible in order for inflation to persist, hence, we can directly solve (\ref{maxwell}) to obtain}
\begin{align}\label{gauge_inisol}
\dot{A}_z = \frac{c_0}{f(\theta_0)^2a}~.
\end{align}
{In order for $R_1,R_2$ to be constant and small throughout inflation, $f(\theta_0) = a^{-2}$ in the classical limit, where here we consider the normalization of the constant. The physical electric field is given by the following in the presence of the dilatonic coupling:} 
\begin{align}
E_{\text{phys}} = -fa^{-1}\frac{dA_z}{dt}~.
\end{align}
{Since $A \propto a^3$ and we require $E_{\text{phys}} = E$, where E is a constant, we would have}
\begin{align}\label{gauge_field_background}
A_z = -\frac{E}{3H}e^{3 H t}~.
\end{align}
{Here, we can also roughly solve for $e(\theta_0)$. In order for $R_2$ in  (\ref{Edensity_ratio}) to have the same order of magnitude as $R_1$, we can set $e(\theta_0) = a^{-2} e_0$ where $e_0$ is the initial amount of charge at the beginning of inflation.} Previous studies \cite{Domenech:2015zzi} mentioned that to generate a large enough electromagnetic field, we would encounter backreaction and strong coupling problems. However, we consider non gauge-invariant coupling to avoid these problems. 
\section{Lagrangian up to the third order}\label{3rdlagrangian}
{We use the ADM formalism to derive the full Lagrangian (\ref{action_f}) up to the third order. We first review the ADM formalism. The full action is given by}
\begin{align}
S = S_g + S_m = \frac{1}{2}\int d^4 x \sqrt{-g} \mathcal{R} + \int d^4 x \sqrt{-g} \mathcal{L}_m~.
\end{align}
{Using the ADM metric,}
\begin{align}
ds^2 = -N^2 dt^2 + h_{ij}(dx^i + N^i dt)(dx^j + N^j dt)~,
\end{align}
{we can decompose the action into}
\begin{align}\label{action_adm}
S = \frac{1}{2}\int d^4 x \sqrt{h} N (R^{(3)} + 2 \mathcal{L}_m) + \frac{1}{2}\int d^4 x \sqrt{h} N^{-1} (E_{ij}E^{ij}-E^2)~. 
\end{align}
{Here, the 3d metric $h_{ij}$ is used to lower the index of $N^i$ and $R^{(3)}$ is the 3d Ricci scalar constructed from $h_{ij}$. We define $E_{ij}$ and $E$ explicitly:}
\begin{align}
E_{ij} &= \frac{1}{2}(\dot{h}_{ij}-\nabla_i N_j - \nabla_j N_i)~, \\
E &= E_{ij}h^{ij}~. 
\end{align}
{We here choose the uniform inflaton gauge,}
\begin{align}
\theta(\mathbf{x},t) = \theta_0(t), \quad \sigma(\mathbf{x},t) &= \sigma_0 + \delta\sigma(\mathbf{x},t), \quad \sigma^*(\mathbf{x},t) = \sigma_0 + \delta\sigma^*(\mathbf{x},t)~,\\
h_{ij}(\mathbf{x},t) &= a^2e^{2 \zeta(\mathbf{x},t)}\delta_{ij}~.
\end{align}
{The constraint equations for the Lagrangian multipliers $N$ and $N_i$ are}
\begin{align}
R^{(3)} + 2\mathcal{L}_m + 2 N \frac{\partial \mathcal{L}_m}{\partial N}-\frac{1}{N^2}(E_{ij}E^{ij}-E^2) = 0~, \\
\nabla_i[N^{-1}(E^{ij}-h^{ij}E)]+N\frac{\partial \mathcal{L}_m}{\partial N_i} = 0~.
\end{align}
{In order to expand the action up to third order in perturbations, we need to solve the Lagrangian multipliers $N$ and $N_i$ up to the first order in perturbations. We then expand,}
\begin{align}\label{adm}
N = 1+\alpha_1, \qquad N_i = \partial_i\psi_1 + \tilde{N}_i^{(1)}~,
\end{align}
{where $\partial_i \tilde{N}_i^{(1)} = 0$. Plugging (\ref{adm}) into (\ref{action_adm}), we can obtain the following expressions:}
\begin{align}\label{adm_sol}
\alpha_1 &= \frac{\dot{\zeta}}{H}, \qquad \tilde{N}_i^{(1)} = 0~,
\end{align}
\begin{align}\label{adm_sol2}
\partial_i \partial_i \psi_1 &= \frac{  -e(\theta_0)A \sigma_0 \text{Im}[ \partial_z \delta\sigma] -2 a^2 R \dot{\theta_0} \text{Re}[ \delta\sigma] +\epsilon a^2 H  \dot{\zeta} -  \partial_i \partial_i \zeta }{  H  }~.
\end{align}
{We plug these solutions back into the action and expand them up to the third order action:}
\begin{align}\label{action_adm_2nd}
\mathcal{L}_2 &= a^3 \epsilon \dot{\zeta}^2 - a \epsilon (\partial_i \zeta)^2 +a^3 \dot{\delta\sigma} \dot{\delta\sigma^*}+\left(a \partial_i \partial_i \delta\sigma  - \frac{\tilde{R}}{R}(a^3 m^2+a e(\theta_0)^2A^2) \delta\sigma \right)\delta\sigma^*+ 2 a e(\theta_0)A  \text{Im}[\delta\sigma \partial_z \delta\sigma^*] \nonumber\\ 
& + 6 a e(\theta_0) A \sigma_0 \text{Im}[\partial_z \delta\sigma^*]\zeta +\left(-4\frac{a^3R\dot{\theta}_0^2}{H}  \text{Re}\left[\delta\sigma \right]+ \frac{2 a e(\theta_0)  A \sigma_0}{H}\text{Im}[ \partial_z\delta\sigma^*]\right)\dot{\zeta}~,
\end{align}
\begin{align}\label{action_adm_3rd}
\mathcal{L}_3 &= a^3  \dot{\delta\sigma} \dot{\delta\sigma^*} \left(3 \zeta - \frac{\dot{\zeta}}{H}\right)-  a \left(3\zeta + \frac{\dot{\zeta}}{H}\right) \partial_i \delta\sigma \partial_i \delta\sigma^* + \frac{H\zeta +\dot{\zeta}}{2 a H^3}\left(- H^2 (\partial_i \partial_j \psi_1)^2 - a^2( m^2\sigma_0^2-\dot{\theta}_0^2 \sigma_0 R)  \text{Im}[\partial_z \delta\sigma]^2  +(\partial_i\partial_i \zeta)^2\right) \nonumber\\
&+ \frac{(\partial_j \psi_1)^2 \partial_i\partial_i\zeta}{a}+a \zeta(\partial_i \zeta)^2 + \epsilon \left(3 a^3 \zeta \dot{\zeta}^2 - \frac{a^3 \dot{\zeta}^3}{H}-\frac{a \dot{\zeta}}{H^2}(H \zeta+\dot{\zeta})\partial_i\partial_i\zeta \right)+\frac{a^3 \dot{\theta}^2_0 \dot{\zeta}(2 \sigma_0 R \dot{\zeta}\text{Re}[\delta \sigma]-H(\sigma_0 + R)\delta\sigma \delta\sigma^*)}{H^2 \sigma_0}+\cdots
\end{align}
{We like to make a comment on (\ref{action_adm_2nd}). We can use it to derive the equation of motion for the perturbations $\zeta$, $\delta\sigma$ and $\delta\sigma^*$ while neglecting the interacting terms that are shown in the second line of (\ref{action_adm_2nd}) and taking $\sigma_0 \ll \tilde{R}$. We here have neglected the higher order terms ($\epsilon^2, A, A^2, A^4$) in expressing the third order action (\ref{action_adm_3rd}). We also can see that in the absence of the electric field, (\ref{adm_sol},\ref{adm_sol2}, \ref{action_adm_2nd}, \ref{action_adm_3rd}) reduced to the expression similar to the ones of quasi-single field inflation \cite{Chen:2009zp}. } 
\section{Angular Power Spectrum due to Vector Field Perturbations}\label{powerspectrum_vfield}
{Here we like to explicitly show the contribution of the vector field perturbations on the angular dependence of the power spectrum. We consider the similar action as (\ref{action_f}) except that we consider the temporal ($A_0$) component and the spacial ($\mathbf{A}$) component of the vector field $A_\mu$. We also consider $\sigma_0 = $ constant, $f = a^{-2}$ and $e(\theta_0) = e_0 a^{-2}$. We obtain the following equations of motion,}
\begin{align}
\nabla \cdot \dot{\mathbf{A}} - \nabla^2 A_0 + 2 a^2 e_0^2 \sigma_0^2 A_0 &= 0 ~,\label{eqn_gaugeeom_1}\\
\ddot{\mathbf{A}} -3 H \dot{\mathbf{A}} - \frac{\nabla^2 \mathbf{A}}{a^2} +2  e_0^2 \sigma_0^2 \mathbf{A} &=  -2 H \nabla A_0~.\label{eqn_gaugeeom_2}
\end{align}
{Here we have use the antisymmetry property of $F^{\mu\nu}$ ($\partial_\mu \partial_\nu F^{\mu\nu} = 0$) in obtaining (\ref{eqn_gaugeeom_2}). We explicitly show the second order Lagrangian that is contributed by the vector field perturbations,}
\begin{align}
\mathcal{L}_{\delta A} &= \frac{1}{2}a f^2 (\partial_i \delta A_0)^2 - a f^2 \nabla \delta A_0 \cdot \dot{\delta \mathbf{A}} + \frac{1}{2}a^3 m_A^2 \delta A_0^2+\frac{1}{2}a f^2  \dot{\delta\mathbf{A}}^2 -\frac{1}{2}a m_A^2 \delta \mathbf{A}^2 -\frac{f^2}{2a}(\nabla \times \delta\mathbf{A} )^2 - a f^2 (\nabla \delta A_0)\cdot \dot{\delta \mathbf{A}}~,
\end{align}
{where $m_A^2 = 2 e(\theta_0)^2 \sigma_0^2$. Hence, we have}
\begin{align}
\nabla \cdot \dot{\delta \mathbf{A}} + \left(2 a^2 e_0^2 \sigma_0^2 - \nabla^2\right) \delta A_0 = 0~,
\end{align}
{for the equation of motion of the temporal gauge. For the spacial gauge, we like to introduce the physical vector field,}
\begin{align}
\delta \mathbf{W} = \frac{f \delta \mathbf{A}}{a}~.
\end{align}
{We obtain the following second order Lagrangian for the spacial gauge}
\begin{align}
\mathcal{L}_{\delta \mathbf{ W}} &= \frac{1}{2}a f^2 \left(\frac{\dot{a} \delta \delta \mathbf{ W}}{f} + \frac{a \dot{ \mathbf{ W}}}{f} - \frac{a \dot{f} \delta \mathbf{ W}}{f^2}\right)^2 -\frac{1}{2}a^3 \frac{m_A^2}{f^2} \delta \mathbf{ W}^2 -\frac{1}{2}a(\nabla \times \delta \mathbf{ W})^2+ (a\dot{a}f +2 a^2 \dot{f} )(\nabla \delta A_0) \cdot \delta  \mathbf{W} \nonumber\\
&+a^2 f \nabla \dot{\delta A_0} \cdot \delta  \mathbf{W}~. 
\end{align}
{We finally obtain the equation of motion for the spacial gauge, }
\begin{align}\label{Wi_eom}
\ddot{\delta \mathbf{W}} + 3 H \dot{\delta \mathbf{W}} + \left( m_0^2 - \frac{\nabla^2}{a^2}\right)\delta \mathbf{W} = -\frac{2H}{a^3}(\nabla \delta A_0)~,
\end{align}
{where $m_0^2 = 2e_0^2 \sigma_0^2$. To investigate the quantum properties of the vector field perturbations, we introduce the following annihilation/creation operators for each polarisation, }
\begin{align}\label{W_expand}
\delta \mathbf{W}(t,\mathbf{x}) &= \sum_{\lambda = L,R,\parallel}\int \frac{d^3 k}{(2\pi)^3}\left(e_\lambda (\hat{\mathbf{k}}) \hat{d}_\lambda (\mathbf{k})w_\lambda(t,k)e^{i \mathbf{k}\cdot\mathbf{x}}+e^*_\lambda (\hat{\mathbf{k}}) \hat{d}^\dagger_\lambda (\mathbf{k})w^*_\lambda(t,k)e^{-i \mathbf{k}\cdot\mathbf{x}}\right)~,
\end{align}
{where $L,R,\parallel$ denotes left and right transverse and longitudinal polarisations respectively. The polarisation vectors are}
\begin{align}
e_L = \frac{1}{\sqrt{2}}(1,i,0), \qquad e_R = \frac{1}{\sqrt{2}}(1,-i,0), \qquad e_\parallel = (0,0,1)~, 
\end{align}
{and $\hat{d}_\lambda,\hat{d}^\dagger_\lambda$ satisfy the usual commutation relations}
\begin{align}
[\hat{d}_\alpha(\mathbf{k}), \hat{d}_\beta^\dagger(\mathbf{k'})] = (2\pi)^3\delta(\mathbf{k}-\mathbf{k'})\delta_{\alpha\beta}~.
\end{align}
{We introduce the expansion (\ref{W_expand}) to (\ref{Wi_eom}), we finally get the equations of motion for the transverse and longitudinal mode functions}
\begin{align}
\ddot{w}_{L,R}+3H\dot{w}_{L,R} + \left(m_0^2 +\frac{k^2}{a^2}\right)w_{L,R} &= 0~, \label{w_modefunctions1}\\
\ddot{w}_{\parallel}+\left(3+\frac{8}{1+r^2}\right)H\dot{w}_{\parallel}+\frac{24}{1+r^2}H^2 w_\parallel +\left(m_0^2 +\frac{k^2}{a^2}\right)w_{\parallel} &= 0~,\label{w_modefunctions2}
\end{align}
{where $r^2 = a^2m_0^2/k^2$. If we consider (\ref{w_modefunctions1},\ref{w_modefunctions2}) to be expressed in terms of conformal time and $r^2 \gg 1$, }
\begin{align}
w''_{L,R}-\frac{2}{\tau}w'_{L,R} + \left(k^2+\frac{m_0^2}{H^2\tau^2} \right)w_{L,R} &= 0~, \label{w_modefunctions1_conformal}\\
w''_{\parallel}-\left(\frac{2}{\tau}+8\tau\frac{k^2H^2}{m_0^2}\right)w'_{\parallel} + \left(k^2+24\frac{k^2H^2}{m_0^2}+\frac{m_0^2}{H^2\tau^2} \right)w_{\parallel} &= 0\label{w_modefunctions2_conformal}~.
\end{align}  
{The solution for the mode functions is given by}
\begin{align}\label{w1_mode}
w_{L,R} &= -i e^{i(\nu+1/2)\pi/2}\frac{\sqrt{\pi}}{2}H(-\tau)^{3/2}H^{(1)}_\nu(-k \tau)~,
\end{align}
{and}
\begin{align}\label{w2_mode}
w_\parallel &= e^{i(-\nu+1/2)\pi/2}\frac{\sqrt{\pi}}{2}\frac{2^\nu k^{-\nu} H(-\tau)^{3/2-\nu}}{\Gamma(1-\nu)\text{sin}(\pi \nu)}\, _1F_1\left(-\frac{m_0^2}{16 H^2}-\frac{\nu }{2}-\frac{3}{4};1-\nu
;\frac{4 H^2 k^2 \tau ^2}{m_0^2}\right)+(\nu \rightarrow -\nu)~,
\end{align}
{where $\nu = \sqrt{9/4-m_0^2/H^2}$. The normalization of (\ref{w1_mode}) is chosen that it recovers the Bunch Davies vacuum when the momentum $k/a$ is larger than the Hubble constant $H$ and the mass of gauge field $m_0$ \cite{Chen:2009we,Chen:2009zp},}
\begin{align}
w_{L,R} \rightarrow i \frac{H}{\sqrt{2k}}\tau e^{-i k \tau}~.
\end{align}
{This is because (\ref{w1_mode}) is the exact solution of (\ref{w_modefunctions1_conformal}). For (\ref{w2_mode}), we just require it to recover to (\ref{w1_mode}) in the large mass $m_A$ limit. If we consider $r^2\ll1$, the mode function $w_\parallel$ decays in the superhorizon limit. From the above mode functions, we can obtain the power spectra in the superhorizon limit. In our model, we consider $r^2\ll 1$ and hence,}
\begin{align}
\mathcal{P}_{L,R} = (\delta W_{L,R})^2 = \left( \frac{H}{2\pi}\right)^2, \qquad \mathcal{P}_\parallel= (\delta W_\parallel)^2 = 0~.
\end{align}
{As shown in \cite{Dimopoulos:2008yv,Dimopoulos:2009vu, Sanchez:2013zst}, the vector perturbations contribute to the curvature power spectrum through the following term}
\begin{align}
\mathcal{P}_{\zeta_A}(\mathbf{k}) = \frac{4\hat{\Omega}_A^2}{9W^2}\left(\mathcal{P}_{L,R}+(\mathcal{P}_\parallel-\mathcal{P}_{L,R})(\hat{\mathbf{W}}\cdot\hat{\mathbf{k}})^2\right)~,
\end{align}
{where $\hat{\mathbf{W}} = \mathbf{W}/W$, $W = |\mathbf{W}|$ and $\hat{\Omega}_A = 3\Omega_A/(4-\Omega_A) \sim \Omega_A = \rho_A/\rho$, where $\rho_A$ is the density of the vector field. We can see that in our model, $\mathcal{P}_{\zeta_A}$ is not isotropic as we are considering the vector field to have a weak mass $m_0$. }
\section{Analytical Expression for the Power Spectrum}\label{analyps}
\begin{figure}[htbp]
	\centering
	\includegraphics[width=12cm]{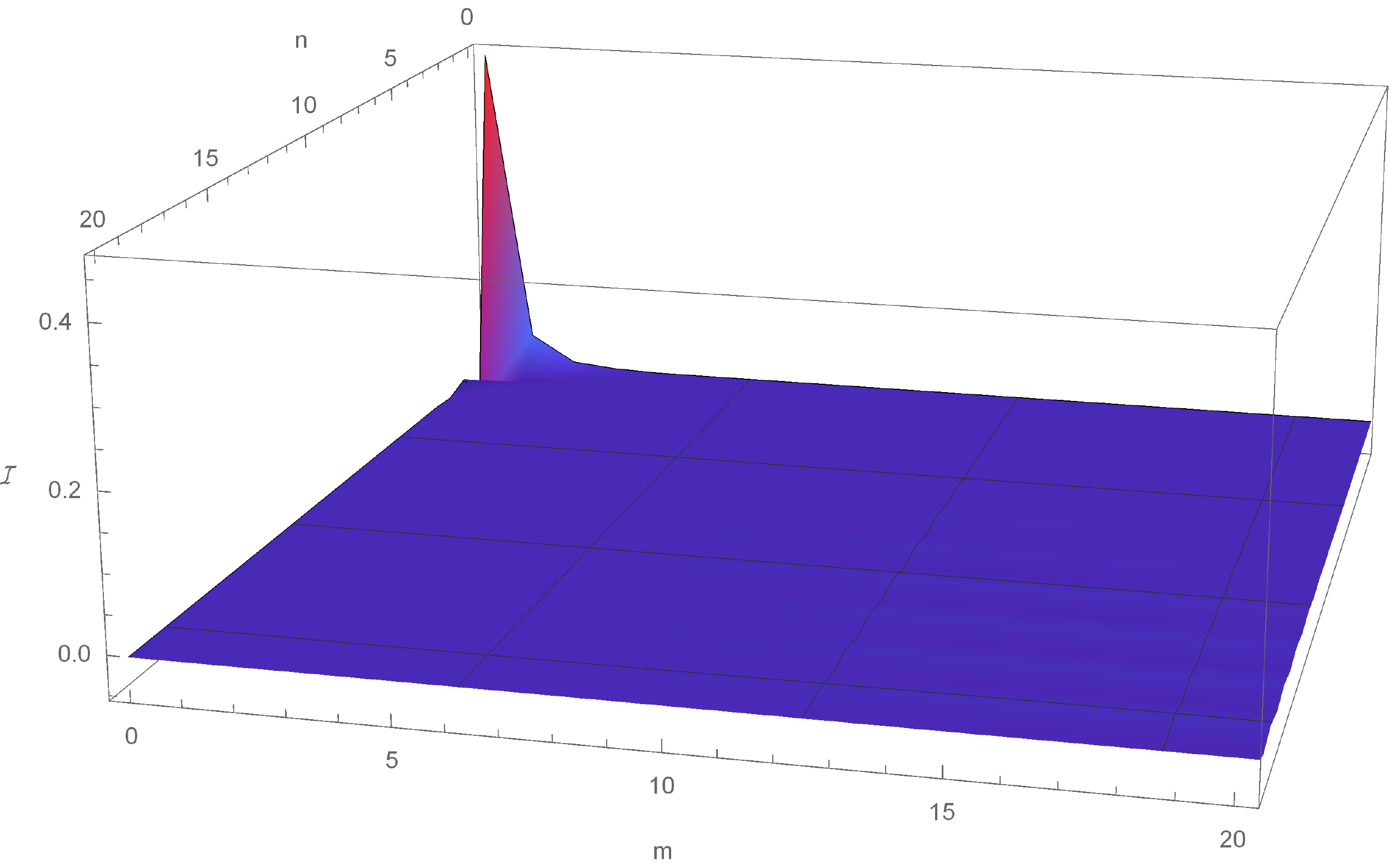}
	\caption{{By setting $\kappa = i$ , $\mu = 5 i$ , we can see that the leading order term in $\mathcal{I}_2$ is given by the $m = 0 , n = 0$ term. This further simplifies our calculation for the power spectrum. }}
	\label{leading}
\end{figure}

{In this section, we compute the analytical expression for the power spectrum by series expansion. In order to calculate the power spectrum, we like to evaluate}
\begin{align}
\Bigg|\int_0^{\infty}dx_1\frac{e^{i x_1} W_{\kappa,\mu}(- 2ix_1)}{x_1}\Bigg|^2\label{formula45}~,
\end{align}
{and}
\begin{align}
\text{Re}\Bigg[\int_0^\infty dx_2\frac{e^{- i x_2} W_{- \kappa,- \mu}(2 i x_2)}{x_2}\int_0^{x_2} dx_1\frac{e^{- i x_1} W_{\kappa,\mu}(-2 i x_1)}{x_1}\Bigg]\label{formula49}~.
\end{align}
{We can expand the Whittaker W as the following}
\begin{align}\label{whittaker}
W_{\kappa , \mu}(z)& = \frac{\Gamma (-2 \mu) z^{\frac{1}{2}+\mu}}{\Gamma (\frac{1}{2} - \mu - \kappa)} \sum_{m=0}^{\infty}\frac{(-\frac{1}{2})^m}{m !} \,_2F_1\left(- m, \mu - \kappa + \frac{1}{2}; 2 \mu + 1; 2\right) z^m\\\nonumber
& + \frac{\Gamma(2 \mu) z^{\frac{1}{2} - \mu}}{\Gamma (\mu - \kappa + \frac{1}{2})}\sum_{m=0}^{\infty} \frac{(-\frac{1}{2})^m}{m !} \, _2F_1 \left(-m, -\mu - \kappa+ \frac{1}{2};2 \mu + 1;2\right)z^m~.
\end{align}
{The calculation of (\ref{formula45}) is trivial as it is similar to (\ref{meijer}): }
\begin{align}\label{4}
\Bigg|\int_0^{\infty}dx_1\frac{e^{i x_1} W_{\kappa,\mu}(- 2ix_1)}{x_1}\Bigg|^2 = \Bigg|\frac{\Gamma(\frac{1}{2} - \mu) \Gamma (\frac{1}{2} + \mu)}{\Gamma (1 - \kappa)}\Bigg|^2~ .
\end{align}
{We calculate the first layer of (\ref{formula49})}
\begin{align}
\mathcal{I}_1 = \int_0^{x_2} dx_1\frac{e^{- i x_1} W_{\kappa,\mu}(-2 i x_1)}{x_1}\label{equi}~,
\end{align}
\begin{align}
\mathcal{I}_1 & = \frac{\Gamma (-2 \mu)}{\Gamma (\frac{1}{2} - \mu - \kappa)} \sum_{m=0}^{\infty}\frac{(-\frac{1}{2})^m (-2)^{\frac{1}{2}+\mu+m}}{m !} \,_2F_1\left(- m, \mu - \kappa + \frac{1}{2}; 2 \mu + 1; 2\right) \int_0^{x_2} d ix_1 e^{- i x_1} (i x_1)^{-\frac{1}{2}+\mu+m}\\\nonumber
& + \frac{\Gamma(2 \mu)}{\Gamma (\mu - \kappa + \frac{1}{2})}\sum_{m=0}^{\infty} \frac{(-\frac{1}{2})^m (-2)^{\frac{1}{2} - \mu+m}}{m !} \, _2F_1 \left(-m, -\mu - \kappa+ \frac{1}{2};2 \mu + 1;2\right)\int_0^{x_2} d ix_1 e^{- i x_1} (i x_1)^{-\frac{1}{2} - \mu + m}\\\nonumber
 & = \frac{\Gamma (-2 \mu)}{\Gamma (\frac{1}{2} - \mu - \kappa)} \sum_{m=0}^{\infty}\frac{(-\frac{1}{2})^m (-2)^{\frac{1}{2}+\mu+m}}{m !} \,_2F_1\left(- m, \mu - \kappa + \frac{1}{2}; 2 \mu + 1; 2\right) \gamma (\frac{1}{2} + \mu + m , i x_2)\\\nonumber
& + \frac{\Gamma(2 \mu)}{\Gamma (\mu - \kappa + \frac{1}{2})}\sum_{m=0}^{\infty} \frac{(-\frac{1}{2})^m (-2)^{\frac{1}{2} - \mu+m}}{m !} \, _2F_1 \left(-m, -\mu - \kappa+ \frac{1}{2};2 \mu + 1;2\right) \gamma (\frac{1}{2} -\mu + m, i x_2)~.\\\nonumber
\end{align}
{Here, we have the following formula to expand the incomplete gamma function}
\begin{align}
\gamma(s,x) = \int_0^x t^{s - 1} e^{- t} dt = x^s \sum_{j = 0}^{\infty} \frac{(- x)^j}{j ! (s + j)}~.
\end{align}
{The result of (\ref{equi}) is of the following}
\begin{align}
\mathcal{I}_1 & = \frac{\Gamma (-2 \mu)}{\Gamma (\frac{1}{2} - \mu - \kappa)} \sum_{m=0}^{\infty}\frac{(-\frac{1}{2})^m (-2)^{\frac{1}{2}+\mu+m}}{m !} \,_2F_1\left(- m, \mu - \kappa + \frac{1}{2}; 2 \mu + 1; 2\right) \sum_{j = 0}^{\infty}(-1)^j \frac{(i x_2)^{\frac{1}{2} + \mu +m +j}}{j ! (\frac{1}{2} + \mu +m +j)}\\\nonumber
& + \frac{\Gamma(2 \mu)}{\Gamma (\mu - \kappa + \frac{1}{2})}\sum_{m=0}^{\infty} \frac{(-\frac{1}{2})^m (-2)^{\frac{1}{2} - \mu+m}}{m !} \, _2F_1 \left(-m, -\mu - \kappa+ \frac{1}{2};2 \mu + 1;2\right)\sum_{j = 0}^{\infty}(-1)^j \frac{(i x_2)^{\frac{1}{2} - \mu +m +j}}{j ! (\frac{1}{2} - \mu +m +j)}~.\\\nonumber
\end{align}
{We perform the following calculation}
\begin{align}
\mathcal{I}_2 = \int_0^{\infty} d x_2 \frac{e^{- i x_2} W_{-\kappa , - \mu}(2 i x_2)}{x_2} \mathcal{I}_1~,
\end{align}
{Here, we use (\ref{whittaker}) to expand $W_{-\kappa , - \mu}(2 i x_2)$ and $\mathcal{I}_2$ can be expressed as the following}
\begin{align}
\mathcal{I}_2 &= \frac{\Gamma (2 \mu)}{\Gamma (\frac{1}{2} + \mu + \kappa)} \sum_{n=0}^{\infty}\frac{(-\frac{1}{2})^n 2^{\frac{1}{2}-\mu+n}}{n !} \,_2F_1\left(- n, -\mu + \kappa + \frac{1}{2}; 1 - 2 \mu ; 2\right) \int_0^{\infty} d ix_2 e^{- ix_2} (i x_2)^{\frac{1}{2}-\mu + n - 1}\mathcal{I}_1\\
& + \frac{\Gamma(-2 \mu)}{\Gamma (-\mu + \kappa + \frac{1}{2})}\sum_{n=0}^{\infty} \frac{(-\frac{1}{2})^n 2^{\frac{1}{2} + \mu + n}}{n !} \, _2F_1 \left(-n, \mu + \kappa+ \frac{1}{2};1 - 2 \mu;2\right) \int_0^{\infty} d ix_2 e^{- i x_2} (i x_2)^{\frac{1}{2} + \mu + n - 1}\mathcal{I}_1\nonumber\\
&= \frac{\Gamma (2 \mu)}{\Gamma (\frac{1}{2} + \mu + \kappa)} \sum_{n=0}^{\infty}\frac{(-\frac{1}{2})^n 2^{\frac{1}{2}-\mu+n}}{n !} \,_2F_1\left(- n, -\mu + \kappa + \frac{1}{2}; 1 - 2 \mu ; 2\right) \nonumber\\
&\Bigg(\frac{\Gamma (-2 \mu)}{\Gamma (\frac{1}{2} - \mu - \kappa)} \sum_{m=0}^{\infty}\frac{(-\frac{1}{2})^m (-2)^{\frac{1}{2}+\mu+m}}{m !} \,_2F_1\left(- m, \mu - \kappa + \frac{1}{2}; 2 \mu + 1; 2\right) \sum_{j = 0}^{\infty}(-1)^j  \frac{\Gamma(1 +m +n +j)}{j ! (\frac{1}{2} + \mu +m +j)}\nonumber\\
& + \frac{\Gamma(2 \mu)}{\Gamma (\mu - \kappa + \frac{1}{2})}\sum_{m=0}^{\infty} \frac{(-\frac{1}{2})^m (-2)^{\frac{1}{2} - \mu+m}}{m !} \, _2F_1 \left(-m, -\mu - \kappa+ \frac{1}{2};2 \mu + 1;2\right)\sum_{j = 0}^{\infty}(-1)^j  \frac{\Gamma (1 - 2 \mu +m +n +j)}{j ! (\frac{1}{2} - \mu +m +j)}\Bigg)\nonumber\\
& + \frac{\Gamma(-2 \mu)}{\Gamma (-\mu + \kappa + \frac{1}{2})}\sum_{n=0}^{\infty} \frac{(-\frac{1}{2})^n 2^{\frac{1}{2} + \mu + n}}{n !} \, _2F_1 \left(-n, \mu + \kappa+ \frac{1}{2};1 - 2 \mu;2\right)\nonumber\\
&\Bigg(\frac{\Gamma (-2 \mu)}{\Gamma (\frac{1}{2} - \mu - \kappa)} \sum_{m=0}^{\infty}\frac{(-\frac{1}{2})^m (-2)^{\frac{1}{2}+\mu+m}}{m !} \,_2F_1\left(- m, \mu - \kappa + \frac{1}{2}; 2 \mu + 1; 2\right) \sum_{j = 0}^{\infty}(-1)^j \frac{\Gamma (1 + 2\mu +m +n +j)}{j ! (\frac{1}{2} + \mu +m +j)}\nonumber\\
& + \frac{\Gamma(2 \mu)}{\Gamma (\mu - \kappa + \frac{1}{2})}\sum_{m=0}^{\infty} \frac{(-\frac{1}{2})^m (-2)^{\frac{1}{2} - \mu+m}}{m !} \, _2F_1 \left(-m, -\mu - \kappa+ \frac{1}{2};2 \mu + 1;2\right)\sum_{j = 0}^{\infty}(-1)^j \frac{\Gamma (1 +m +n +j)}{j ! (\frac{1}{2} - \mu +m +j)}\Bigg)~.\nonumber
\end{align}
{We then perform the following summation to simplify $\mathcal{I}_2$}
\begin{align}
\sum_{j = 0}^{\infty}(-1)^j  \frac{\Gamma( p + j )}{j ! (q +j)} = \frac{\Gamma (p) \,_2F_1 \left(p,q; q+1;-1\right)}{q}~.
\end{align}
{So, $\mathcal{I}_2$ can be expressed in following form,}
\begin{align}\label{I2}
\mathcal{I}_2&= \frac{\Gamma (2 \mu)}{\Gamma (\frac{1}{2} + \mu + \kappa)} \sum_{n=0}^{\infty}\sum_{m=0}^{\infty}\frac{(-1)^{\frac{1}{2}+ n + 2 m} 2^{1-\mu}}{n !} \,_2F_1\left(- n, -\mu + \kappa + \frac{1}{2}; 1 - 2 \mu ; 2\right) \\
&\Bigg(\frac{\Gamma (-2 \mu)\Gamma (1 +m +n )}{\Gamma (\frac{1}{2} - \mu - \kappa)(\frac{1}{2} + \mu +m)} \frac{(-2)^{\mu}}{m !} \,_2F_1\left(- m, \mu - \kappa + \frac{1}{2}; 2 \mu + 1; 2\right)\,_2F_1 \left(1 +m +n ,\frac{1}{2} + \mu +m ; \frac{3}{2} + \mu +m ;-1\right)+\nonumber\\
& \frac{\Gamma(2 \mu)\Gamma (1 - 2 \mu +m +n)}{\Gamma (\mu - \kappa + \frac{1}{2})(\frac{1}{2} - \mu +m)} \frac{(-2)^{- \mu}}{m !} \, _2F_1 \left(-m, -\mu - \kappa+ \frac{1}{2};2 \mu + 1;2\right)\,_2F_1 \left(1 - 2 \mu +m +n,\frac{1}{2} - \mu +m; \frac{3}{2} - \mu +m;-1\right)\Bigg)\nonumber\\
& + \frac{\Gamma(-2 \mu)}{\Gamma (-\mu + \kappa + \frac{1}{2})}\sum_{n=0}^{\infty} \sum_{m=0}^{\infty}\frac{(-1)^{\frac{1}{2}+ n + 2 m} 2^{1 + \mu}}{n !} \, _2F_1 \left(-n, \mu + \kappa+ \frac{1}{2};1 - 2 \mu;2\right)\nonumber\\
&\Bigg(\frac{\Gamma (-2 \mu)\Gamma (1 + 2\mu +m +n)}{\Gamma (\frac{1}{2} - \mu - \kappa)(\frac{1}{2} + \mu +m)} \frac{(-2)^{\mu}}{m !} \,_2F_1\left(- m, \mu - \kappa + \frac{1}{2}; 2 \mu + 1; 2\right)\,_2F_1 \left(1 + 2\mu +m +n,\frac{1}{2} + \mu +m; \frac{3}{2} + \mu +m;-1\right)\nonumber\\
& + \frac{\Gamma(2 \mu)\Gamma (1 +m +n)}{\Gamma (\mu - \kappa + \frac{1}{2})(\frac{1}{2} - \mu +m)} \frac{(-2)^{- \mu}}{m !} \, _2F_1 \left(-m, -\mu - \kappa+ \frac{1}{2};2 \mu + 1;2\right) \,_2F_1 \left(1 +m +n,\frac{1}{2} - \mu +m; \frac{3}{2} - \mu +m;-1\right)\Bigg)~.\nonumber
\end{align}
{To perform a better analysis on $\mathcal{I}_2$, we divide $\mathcal{I}_2$ into four parts as the following,}
\begin{align}
P_1 &= \frac{\Gamma (2 \mu)\Gamma (-2 \mu)}{\Gamma (\frac{1}{2} + \mu + \kappa)\Gamma (\frac{1}{2} - \mu - \kappa)} \sum_{n=0}^{\infty}\sum_{m=0}^{\infty}\frac{(-1)^{\frac{1}{2}+ \mu+ n + 2 m} 2}{n ! m! (\frac{1}{2} + \mu +m)} \Gamma (1 +m +n ) \\\nonumber
&\,_2F_1\left(- n, -\mu + \kappa + \frac{1}{2}; 1 - 2 \mu ; 2\right) \,_2F_1\left(- m, \mu - \kappa + \frac{1}{2}; 2 \mu + 1; 2\right)\,_2F_1 \left(1 +m +n ,\frac{1}{2} + \mu +m ; \frac{3}{2} + \mu +m ;-1\right)~,\\
P_2 &= \frac{\Gamma (2 \mu)\Gamma(2 \mu)}{\Gamma (\frac{1}{2} + \mu + \kappa)\Gamma (\mu - \kappa + \frac{1}{2})} \sum_{n=0}^{\infty}\sum_{m=0}^{\infty}\frac{(-1)^{\frac{1}{2} - \mu + n + 2 m} 2^{1- 2\mu}}{n ! m ! (\frac{1}{2} - \mu +m)} \Gamma (1 - 2 \mu +m +n)\\\nonumber
& \,_2F_1\left(- n, -\mu + \kappa + \frac{1}{2}; 1 - 2 \mu ; 2\right) \, _2F_1 \left(-m, -\mu - \kappa+ \frac{1}{2};2 \mu + 1;2\right)\,_2F_1 \left(1 - 2 \mu +m +n,\frac{1}{2} - \mu +m; \frac{3}{2} - \mu +m;-1\right)~,\\
P_3 &= \frac{\Gamma(-2 \mu)\Gamma (-2 \mu)}{\Gamma (-\mu + \kappa + \frac{1}{2})\Gamma (\frac{1}{2} - \mu - \kappa)}\sum_{n=0}^{\infty} \sum_{m=0}^{\infty}\frac{(-1)^{\frac{1}{2}+\mu + n + 2 m} 2^{1 + 2 \mu}}{n ! m ! (\frac{1}{2} + \mu +m)} \Gamma (1 + 2\mu +m +n)\\\nonumber
&\, _2F_1 \left(-n, \mu + \kappa+ \frac{1}{2};1 - 2 \mu;2\right) \,_2F_1\left(- m, \mu - \kappa + \frac{1}{2}; 2 \mu + 1; 2\right)\,_2F_1 \left(1 + 2\mu +m +n,\frac{1}{2} + \mu +m; \frac{3}{2} + \mu +m;-1\right)~,\\
P_4 &= \frac{\Gamma(-2 \mu)\Gamma(2 \mu)}{\Gamma (-\mu + \kappa + \frac{1}{2})\Gamma (\mu - \kappa + \frac{1}{2})}\sum_{n=0}^{\infty} \sum_{m=0}^{\infty}\frac{(-1)^{\frac{1}{2} -\mu+ n + 2 m} 2}{n ! m ! (\frac{1}{2} - \mu +m)} \Gamma (1 +m +n)\\\nonumber
& \, _2F_1 \left(-n, \mu + \kappa+ \frac{1}{2};1 - 2 \mu;2\right) \, _2F_1\left(-m, -\mu - \kappa+ \frac{1}{2};2 \mu + 1;2\right) \,_2F_1 \left(1 +m +n,\frac{1}{2} - \mu +m; \frac{3}{2} - \mu +m;-1\right)~.\\\nonumber
\end{align}
We can see that each $\Gamma (\pm n\mu + \cdots) $  function induces a factor of $e^{-n |\mu|}$ and any hypergeometric function $\, _2F_1$ does not induce such a factor. {We can plot each term of the series expansion of \eqref{I2} in FIG. \ref{leading}. We can see that the main contribution comes from the $m=0,n=0$ term of $P_1$. This is because the coefficient of $P_1$ of the summation series dominates the other 3 terms $(P_2,P_3,P_4)$ and for large $m$ and $n$, the terms are exponentially suppressed. We then can make the following approximation to $\mathcal{I}_2$.}

\begin{align}
\mathcal{I}_2 &\approx  \frac{\Gamma (2 \mu)\Gamma (-2 \mu)}{\Gamma (\frac{1}{2} + \mu + \kappa)\Gamma (\frac{1}{2} - \mu - \kappa)} \frac{(-1)^{\frac{1}{2}+ \mu} 2}{ (\frac{1}{2} + \mu )} \\\nonumber
&\,_2F_1\left(0 , -\mu + \kappa + \frac{1}{2}; 1 - 2 \mu ; 2\right) \,_2F_1\left( 0 , \mu - \kappa + \frac{1}{2}; 2 \mu + 1; 2\right)\,_2F_1 \left(1 ,\frac{1}{2} + \mu ; \frac{3}{2} + \mu ;-1\right)\\\nonumber
& = \frac{i e^{-i \pi  \mu }}{2 \mu } \csc (2 \pi  \mu ) \cos (\pi  (\kappa+\mu )) \left(\psi ^{(0)}\left(\frac{\mu}{2}+\frac{1}{4}\right)-\psi ^{(0)}\left(\frac{\mu }{2}+\frac{3}{4}\right)\right)~.
\end{align}
We can easily generalized the analytic expression of the power spectrum to the case large mass limit. Similar results can be obtained in \cite{Tolley:2009fg,Achucarro:2010jv,Achucarro:2012sm,Gwyn:2012mw,Chen:2012ge,Pi:2012gf,An:2017hlx,Tong:2017iat,Iyer:2017qzw}. 
\section{Bispectrum in the Large Mass Limit }\label{largemassbispectrum}
In this section, we derive the analytical expression for the bispectrum in {the large mass limit}. 
The large mass limit of the bispectrum in the quasi-single field model with neutral scalar particles is obtained in \cite{Gong:2013sma}. After integrating out the massive field, an equilateral non-Gaussianity is obtained. This part contains no clock signal, however, it is the dominant contribution in the large mass limit since it is only suppressed by $1/|\mu|^2$ whereas the clock signal is supressed by $\exp(-\pi|\mu|)$ in the large mass limit. In this section, we like to derive the bispectrum by integrating out the {massive charged} scalar particles. 

First, we calculate the second term of the  bispectrum, we can write it as
\begin{align}\nonumber
\langle \zeta_{\mathbf k_1} \zeta_{\mathbf k_2}\zeta_{\mathbf k_3}\rangle'^{(2)} &= 2  c_2 c^*_3 \frac{H^3 e^{i\pi\kappa}}{128 \epsilon^3 k_1 k_2 k^4_3 M_{\rm pl}^6}   {\rm Re} \bigg[ \int_{0}^{\infty} dx_1  \frac{e^{-i x_1} W_{\kappa,\mu}(-2ix_1)}{x_1} \int_{x_1}^{\infty} dx_2 x_2  e^{-i  \frac{k_1+k_2}{k_3} x_2} W_{-\kappa,-\mu}(2i x_2)  \bigg]\\
&  + (\kappa\rightarrow - \kappa ,c_2\rightarrow c_2^*,c_3^*\rightarrow c_3)~. 
\end{align}
We expand the Whittaker function in the limit $x\rightarrow 0$, the first layer of integral is evaluated to
\begin{align}
\int_{x_1}^{\infty} dx_2 x_2  e^{-i  \frac{k_1+k_2}{k_3} x_2} \frac{(2 i x_2)^{1/2-\mu}\Gamma (2\mu)}{\Gamma \left(\kappa +\mu +\frac{1}{2}\right)} \sim - (2 i)^{\frac{1}{2}-\mu}\frac{\Gamma (2\mu)}{\Gamma (\kappa+\mu+\frac{1}{2})}\frac{(x_{1})^{\frac{5}{2}-\mu}}{-\mu + \frac{5}{2}}e^{-i  \frac{k_1+k_2}{k_3} x_1}~.
\end{align}
Inserting it into the second integral
\begin{align}\nonumber
&\int_{0}^{\infty} dx_1 \frac{e^{-i x_1}}{x_1}\frac{(-2 i x_1)^{1/2+\mu}\Gamma (-2\mu)}{\Gamma \left(-\kappa -\mu +\frac{1}{2}\right)}(-1) (2 i)^{\frac{1}{2}-\mu}\frac{\Gamma (2\mu)}{\Gamma (\kappa+\mu+\frac{1}{2})}\frac{( x_{1})^{\frac{5}{2}-\mu}}{-\mu + \frac{5}{2}}e^{-i  \frac{k_1+k_2}{k_3} x_1}\\\nonumber
&= (2i)(-1)^{\frac{3}{2}+\mu}\frac{\Gamma(2\mu)\Gamma(-2\mu)}{\Gamma(\kappa+\mu+\frac{1}{2})\Gamma(-\kappa-\mu+\frac{1}{2})(-\mu+\frac{5}{2})} \int_0^{\infty}dx_1 x_1^2 e^{-i x_1 \frac{k_1 +k_2 +k_3}{k_3}}\\
&=2 (-1)^{\frac{1}{2}+\mu} \frac{\csc (2\pi \mu)\cos (\pi(\kappa + \mu))}{\mu(\mu-\frac{5}{2})}\bigg(\frac{k_3}{k_1 +k_2 +k_3}\bigg)^3~.
\end{align}
Hence, the second term of the bispectrum in the large mass limit can be expressed as
\begin{align}
\langle \zeta_{\mathbf k_1} \zeta_{\mathbf k_2}\zeta_{\mathbf k_3} \rangle'^{(2)} &= 2 c_2 c^*_3 \frac{H^3 e^{i \pi \kappa}}{128 \epsilon^3 M_{\rm pl}^6 k_1 k_2 k_3 (k_1 +k_2 +k_3 )^3}{\rm Re}\left[2 (-1)^{\frac{1}{2}+\mu}\frac{\csc (2\pi \mu)\cos (\pi(\kappa + \mu))}{\mu(\mu-\frac{5}{2})}\right] \nonumber\\
&+(\kappa \rightarrow -\kappa ,c_2\rightarrow c_2^*,c_3^*\rightarrow c_3)~.
\end{align}
We can get the third term of the bispectrum in the large mass limit in the similar way.
\begin{align}
\langle \zeta_{\mathbf k_1} \zeta_{\mathbf k_2}\zeta_{\mathbf k_3} \rangle'^{(3)} &= 2 c_2 c^*_3\frac{H^3 e^{i \pi \kappa}}{128 \epsilon^3 M_{\rm pl}^6 k_1 k_2 k_3 (k_1 +k_2 +k_3 )^3}{\rm Re}\left[2 (-1)^{\frac{1}{2}-\mu}\frac{\csc (2\pi \mu)\cos (\pi(\kappa + \mu))}{\mu(\mu-\frac{1}{2})}\right]\nonumber\\
&+(\kappa \rightarrow -\kappa ,c_2\rightarrow c_2^*,c_3^*\rightarrow c_3)~.
\end{align}

\section{Analytical Expression for the Clock Signal}\label{clock}
In this section, we compute the analytical expression for the clock signal following the standard approach developed in \cite{Arkani-Hamed:2015bza}. In order to calculate the clock signal of primordial bispectrum analytically, we first simplify the propagators $D_{++}$, $D_{+-}$, $D_{-+}$, and $D_{--}$ in the following way. Using the late time behavior \eqref{latetimebehavior} and focusing on the non-local terms 
\begin{align}\nonumber
v_{\mathbf k}(\tau_1) v^*_{\mathbf k} (\tau_2) = \frac{e^{-\pi|\mu|}}{a(\tau_1)a(\tau_2)4 k |\mu|} \bigg(& |\alpha|^2 M_{\kappa,\mu}(2ki\tau_1)(M_{\kappa,\mu}(2ik\tau_2))^*+ \alpha\beta^*  M_{\kappa,\mu}(2ki\tau_1)M_{\kappa,\mu}(2ik\tau_2)\\
&+\alpha^*\beta (M_{\kappa,\mu}(2ki\tau_1))^*(M_{\kappa,\mu}(2ik\tau_2))^*+|\beta|^2 (M_{\kappa,\mu}(2ki\tau_1))^* M_{\kappa,\mu}(2ik\tau_2) \bigg)~, \\\nonumber
v^*_{\mathbf k}(\tau_1) v_{\mathbf k} (\tau_2) = \frac{e^{-\pi|\mu|}}{a(\tau_1)a(\tau_2)4 k |\mu|} \bigg(& |\alpha|^2 (M_{\kappa,\mu}(2ki\tau_1))^* M_{\kappa,\mu}(2ik\tau_2) + \alpha\beta^*  M_{\kappa,\mu}(2ki\tau_1)M_{\kappa,\mu}(2ik\tau_2)\\
&+\alpha^*\beta (M_{\kappa,\mu}(2ki\tau_1))^*(M_{\kappa,\mu}(2ik\tau_2))^*+|\beta|^2  M_{\kappa,\mu}(2ki\tau_1) ( M_{\kappa,\mu}(2ik\tau_2) )^* \bigg)~,
\end{align}
from where we know that only $|\alpha|^2$ and $|\beta|^2$ terms are different. However, the $|\alpha|^2$ and $|\beta|^2$ terms are local, and hence, do not contribute to the clock signal.In order to understand the bispectrum in the squeeze limt ( $\alpha\beta^*$ and $\alpha^*\beta$ ), the four types of propagators $D_{++}$, $D_{+-}$, $D_{-+}$ and $D_{--}$, defined in \eqref{propagator1} becomes identical.
\begin{align}\nonumber
&D({\mathbf k},\tau_1,\tau_2)\equiv D_{++}({\mathbf k},\tau_1,\tau_2)=D_{-+}({\mathbf k},\tau_1,\tau_2)=D_{+-}({\mathbf k},\tau_1,\tau_2)=D_{--}({\mathbf k},\tau_1,\tau_2) \\
&=\frac{e^{-\pi|\mu|}}{a(\tau_1)a(\tau_2)4 k |\mu|} \bigg( \alpha\beta^*  M_{\kappa,\mu}(2ki\tau_1)M_{\kappa,\mu}(2ik\tau_2) +\alpha^*\beta (M_{\kappa,\mu}(2ki\tau_1))^*(M_{\kappa,\mu}(2ik\tau_2))^*  \bigg)~.
\end{align}
The squeezed limit bispectrum can be calculated for all orders in the $(k_1/k_3)$ expansion. However, for simplicity, we focus on the first order $(k_1/k_3)$ expansion, where the Whittacker M is expanded as
\begin{align}
M_{\kappa,\mu} (z) = z^{\mu+1/2} + \mathcal O(z^{\mu+3/2})~.
\end{align}
Taking the effective mass to be {$e_0^2E^2/H^4+m^2/H^2>9/4$.}
The charged massive scalar propagator becomes 
\begin{align}\nonumber
D({\mathbf k},\tau_1, \tau_2) &\equiv D_{++}({\mathbf k},\tau_1,\tau_2)=D_{-+}({\mathbf k},\tau_1,\tau_2)=D_{+-}({\mathbf k},\tau_1,\tau_2)=D_{--}({\mathbf k},\tau_1,\tau_2) \\
&= \frac{e^{-\pi|\mu|}}{a(\tau_1)a(\tau_2)4k|\mu|} \bigg(\alpha^*\beta (-4k^2\tau_1\tau_2)^{-\mu+1/2}   + {\rm c.c}  \bigg)~.
\end{align} 
The bispectrum is further simplified to
\begin{align}\nonumber
\langle \zeta_{\mathbf k_1} \zeta_{\mathbf k_2}  \zeta_{\mathbf k_3}  \rangle' = 2 c_2 c_3^* {\rm Re} \Bigg[\int_{-\infty}^{0}   \int_{-\infty}^{0}   &\frac{d\tau_1}{(-H\tau_1)^3} \frac{d\tau_2}{(-H\tau_2)^2}  [\partial_{\tau_1}G_{++}({\mathbf k_3},\tau_1,0)-\partial_{\tau_1}G_{-+}({\mathbf k_3},\tau_1,0)]D({\mathbf k_3},\tau_1,\tau_2)\\
& 
\partial_{\tau_2}G_{++}({\mathbf k_1},\tau_2,0)\partial_{\tau_2}G_{++}({\mathbf k_2},\tau_2,0) \Bigg]+ (\kappa \rightarrow -\kappa,c_2\rightarrow c_2^*,c_3^*\rightarrow c_3 )~.
\end{align}
After evaluation, the bispectrum becomes
\begin{align}
\langle \zeta_{\mathbf k_1} \zeta_{\mathbf k_2}  \zeta_{\mathbf k_3}  \rangle'   =  c_2 c^*_3 {\rm Re} \Bigg[  f(\mu,\kappa)   \frac{
2^{2\mu}H^3 }{16k_1k_2k_3^{3/2}k_{12}^{5/2}M_{\rm pl}^6 \epsilon^3}   \bigg(\frac{k_1+k_2}{k_{3}}\bigg)^{-\mu}  \Bigg] +(\kappa \rightarrow -\kappa,c_2\rightarrow c_2^*,c_3^*\rightarrow c_3 )~,
\end{align}
where the prefactor is given by 
\begin{align}
f(\mu,\kappa)\equiv   e^{i \pi \kappa }  \frac{\Gamma(-2\mu)^2\Gamma\left(\frac{1}{2}+\mu\right)\Gamma\left(\frac{5}{2}+\mu\right)}{\Gamma(\frac{1}{2}-\mu-\kappa)\Gamma(\frac{1}{2}+\mu-\kappa)}\left(1+ \text{sin} \left(\pi \mu \right)\right)~. 
\end{align}
The shape function is
\begin{align}
S(k_1,k_2,k_3) = \frac{c_2c_3^*}{H M_{\rm pl}^2 \epsilon} {\rm Re} \bigg[f(\mu,\kappa) 
\frac{
	2^{2\mu-2}k_1k_2 k_3^{1/2}}{k_{12}^{5/2} }   \bigg(\frac{k_1+k_2}{k_{3}}\bigg)^{-\mu}  \bigg]+ (\kappa \rightarrow -\kappa,c_2\rightarrow c_2^*,c_3^*\rightarrow c_3 )~.
\end{align}

\end{document}